\documentclass[fleqn,usenatbib]{mnras}
\usepackage{newtxtext,newtxmath}

\usepackage[utf8]{inputenc}
\usepackage[T1]{fontenc}
\usepackage{ae,aecompl}
\usepackage{graphicx}	
\usepackage{amsmath}	
\usepackage{amssymb}	
\usepackage[normalem]{ulem}
\usepackage{multicol}
\usepackage{pdflscape}

\newcommand{\noop}[1]{}

\usepackage{longtable}
\usepackage{array}
\usepackage{rotating}
\newcommand{\PreserveBackslash}[1]{\let\temp=\\#1\let\\=\temp}
\newcolumntype{C}[1]{>{\PreserveBackslash\centering}p{#1}}
\newcolumntype{R}[1]{>{\PreserveBackslash\raggedleft}p{#1}}
\newcolumntype{L}[1]{>{\PreserveBackslash\raggedright}p{#1}}

\title[HdC Spectral Classification]{A Spectral Classification System for Hydrogen-deficient Carbon Stars} 
\author[C. L. Crawford et al.]{
Courtney L. Crawford$^{1,2}$\thanks{Email: courtney.crawford@sydney.edu.au},
Patrick Tisserand$^{3}$,
Geoffrey C. Clayton$^{2}$,
Jamie Soon$^{4}$,
\newauthor Mike Bessell$^{4}$,
Peter Wood$^{4}$,
D. A. Garc\'{i}a-Hern\'{a}ndez${^{5,6}}$,
Ashley J. Ruiter$^{7}$,
\newauthor Ivo R. Seitenzahl$^{7}$
\\
$^{1}$Sydney Institute for Astronomy (SIfA), School of Physics, University of Sydney, NSW 2006, Australia\\
$^{2}$Dept. of Physics \& Astronomy, Louisiana State University, Baton Rouge, LA, 70803, USA\\
$^{3}$Sorbonne Universit\'es, UPMC Univ. Paris 6 et CNRS, UMR 7095, Institut d’Astrophysique de Paris, IAP, 75014 Paris, France\\
$^{4}$Research School of Astronomy and Astrophysics, Australian National University, Cotter Rd, Weston Creek ACT 2611, Australia\\
$^{5}$Instituto de Astrof\'{i}sica de Canarias (IAC), E-38205 La Laguna, Tenerife, Spain\\
$^{6}$Universidad de La Laguna (ULL), Departamento de Astrof\'{i}sica, E-38206 La Laguna, Tenerife, Spain\\
$^{7}$School of Science, University of New South Wales Canberra, The Australian Defence Force Academy, 2600 ACT, Canberra, Australia
}
\date{Accepted --. Received --; in original form October 10, 2022}
\pubyear{2022}
\begin{document}
\label{firstpage}
\pagerange{\pageref{firstpage}--\pageref{lastpage}}
\maketitle

\begin{abstract}
Stellar spectral classification has been highly useful in the study of stars. While there is a currently accepted spectral classification system for carbon stars, the subset of Hydrogen-deficient Carbon (HdC) stars has not been well described by such a system, due predominantly to their rarity and their variability. Here we present the first system for the classification of HdCs based on their spectra, which is made wholly on their observable appearance. We use a combination of dimensionality reduction and clustering algorithms with human classification to create such a system with eight total classes corresponding to temperature, and an additional second axis corresponding to the carbon molecular band strength. We classify over half of the known sample of HdC stars using this, and roughly calibrate the temperatures of each class using their colors. Additionally, we express trends in the occurrence of certain spectral peculiarities such as the presence of Hydrogen and Lithium lines. We also present three previously unpublished spectra, report the discovery of two new Galactic dustless HdC (dLHdC) stars and additionally discuss one especially unique star that appears to border between the hottest HdCs and the coolest Extreme Helium (EHe) stars.
\end{abstract}


\section{Introduction}

Taxonomy is used across all of the sciences, and allows the direct communication of characteristic traits by simply providing an object's class. The goal of stellar spectroscopic taxonomy is no different-- one would like to communicate stellar parameters simply. The categorization of stars into classes began with \citet{Secchi1868} which split 4,000 spectra of stars into 5 separate classes, one of which (class IV) we now know corresponds to carbon stars. This system was succeeded by the Draper system \citep{Pickering1980_HDclass}, which used letters A through Q to describe the spectral patterns. \citet{Cannon1918_harvardclass} later created the Harvard system which rearranged the classes used in the Draper system, forming the typical order we know of today-- O, B, A, F, G, K, and M. They noted the spectral peculiarities of some stars such as R Coronae Borealis (R CrB), which was later confirmed by \citet{Bidelman1953_hydrogendeficient} to be hydrogen-deficient. The final adjustment was made in \citet{Morgan1943_MKKsystem} which moved the classification system into two dimensions, forming the Yerkes (or MKK or MK) system. \citet{Morgan1984_mkprocess} coined the term ``MK Process'' to describe the philosophy by which one should create, use, and maintain a stellar classification system similar to the Yerkes system. One of the main tenets is that the system is defined solely by its own standard spectra, with no comments on the underlying physics which creates the differences in spectra. 

The MK process has been used to create spectral classification systems for many other classes of stars, one of which being the current carbon star classification system \citep{Keenan1993_carbonclass,Barnbaum1996_carbon}. Carbon stars are classified under this system into one of five types: C-R and C-N being the most common, followed by C-J and C-H, and the rarest being C-Hd. These types are arranged such that a star with class C-R4 and one with class C-N4 have approximately the same temperature. Each star is also given a C$_2$ index that corresponds to the strength of the C$_2$ Swan bands. Of the carbon stars, C-N stars are the most common type, characterized by extreme redness, weak carbon isotopic bands, and enhanced \textit{s}-process features. C-R stars tend to be warmer, have intermediate strength isotopic bands, and nearly solar \textit{s}-process features. C-J stars are stars with unusually strong $^{13}$C isotopic bands, and C-H stars are stars with exceptionally strong CH bands (the G-band at 4300 \text{\AA}) . C-Hd stars are the rarest class, with only three stars classified as this type in \citet{Barnbaum1996_carbon}. These three stars are examples of Hydrogen-deficient Carbon (HdC) stars. 

HdC stars are a set of carbon stars which show exceptionally weak or no hydrogen features in their spectra, as well as a severe depletion in $^{13}$C and strong $^{12}$C$^{18}$O bands in the infrared \citep{Warner1967_hdcs,Clayton1996_review,Clayton2007_O18,Clayton2012_review}. They come in two varieties: the dustless HdC (dLHdC) stars and the R Coronae Borealis (RCB) variables. Note that the three stars classified in \citet{Barnbaum1996_carbon} are all dLHdCs. RCB stars have typically been avoided by classifiers due to their unique variability caused by the formation of circumstellar dust clouds at irregular intervals and their enigmatic pulsations \citep{Saio2008_strangemodercb}. The number of known HdC stars has skyrocketed in recent years due to advancements in selecting candidates based on infrared colors \citep{Tisserand2012_IRcolors,Tisserand2020_plethora,Tisserand2022_dlhdcdiscovery}. There are currently 162 RCB stars and 34 dLHdC stars known in the Milky Way and the Magellanic Clouds, making them an extremely rare class of star. HdC stars have higher temperatures (5000-8000 K by some estimates) than the carbon star population (2500-6000 K), and thus it is not straightforward to classify HdC stars using the existing carbon star criteria \citep{Asplund2000,Keenan1993_carbonclass}. 
We therefore present a new classification system specifically for the HdC class of stars.

Additionally, we have begun to find spectroscopic differences between RCBs and dLHdCs. dLHdCs are known to have stronger $^{18}$O in the IR \citep{Karambelkar2022_oxygen18}, and are more likely to be Sr-rich than RCBs \citep{Crawford2022_strontium}. Additionally, dLHdCs appear to be more likely to have visible Balmer lines than RCBs, and may have weaker CN bands \citep{Tisserand2022_dlhdcdiscovery}. After recent analyses by \citet{Tisserand2022_dlhdcdiscovery} and \citet{Karambelkar2022_oxygen18}, it seems possible that dLHdCs and RCBs are formed by different populations of white dwarf (WD) mergers, as dLHdCs occupy a slightly different space in the HR diagram, although the dLHdCs could also be RCBs in an ``off'' state, where they are not producing dust for long periods of time. 
See \citet{Tisserand2022_dlhdcdiscovery} for a detailed discussion of dLHdCs compared to RCBs. For the purposes of this work, we will include both RCB (as close to maximum light as possible) and dLHdC spectra. In doing so, we do not find any new spectroscopic differences at intermediate resolution.

Recently, it has been suggested that the rare class of DY Per type variable stars could be related to the RCB variables \citep{Alksnis1994_dypertheory,Bhowmick2018_coolcousins}. The DY Per variables are characterized by very cold, carbon-rich spectra and by their light curve variations, showing pulsations at maximum light and irregular dust declines. DY Per type declines are much weaker than RCB declines (up to 2 mag vs. up to 8 mag) and appear more symmetrical in nature. RCB declines show a sharp initial decline, followed by a slow rise back to maximum light, whereas the DY Per types have a comparably slow initial decline and a slow return to maximum. DY Per type variables are indeed carbon stars like the RCBs and dLHdCs, however their hydrogen-deficiency is less obvious. As the DY Per types are quite cold ($\sim$3500 K for the prototype, DY Per \citep{Keenan1997_dyper}), the Balmer series is not easily detected and there is not enough spectral flux near the G band at 4300 \text{\AA} to confirm nor deny its existence. However, previous optical and mid-IR Spitzer spectra suggest that the prototype DY Per could indeed be somewhat hydrogen-deficient \citep{Zacs2007_dyper,Yakovina2009_dyper,GarciaHernandez2013_dust}. Therefore, DY Per types are included in the HdC umbrella by definition, however in this work we do not include them as we do not currently have classifiable spectra for these stars (see Section~\ref{subsec:dyper_calib}). We do, however, include comparison between the coolest RCBs and the DY Per type variables whenever possible.

This work outlines the first spectral classification for HdC stars. We describe our observations of the HdC stars in Section~\ref{sec:obs}. In Section~\ref{sec:methods} we outline the methods used to create the classification system, and we list the the specific criteria useful for classification in Section~\ref{sec:criteria}. We discuss a preliminary calibration of the system in Section~\ref{sec:calib} and compare to the DY Per type variables in Section~\ref{subsec:dyper_calib}. We discuss a special case of warm HdC star in Section~\ref{sec:j005}. Finally, we conclude in Section~\ref{sec:conclusions}.

\section{Observations}
\label{sec:obs}

There are 196 total HdC stars known in the Milky Way and the Magellanic Clouds \citep{Tisserand2020_plethora,Tisserand2022_dlhdcdiscovery}. Five of these stars are considered ``hot'' RCBs, which are not considered in this article due to their diversity from the main set of HdCs \citep{Demarco2002_hotrcbs}. We obtained blue and red optical spectra for 144 of these stars using the Wide Field Spectrograph (WiFeS) \citep{Dopita2007_wifes} mounted on the 2.3m telescope of the Australian National University (ANU) at Siding Spring Observatory, using the B3000 and R3000 gratings. Many of these spectra were also used in \citet{Tisserand2020_plethora, Tisserand2022_dlhdcdiscovery}. These spectra span a wavelength range of 3400 to 9600 \text{\AA}, with a two-pixel resolution of about 2 \text{\AA}. For classification purposes, it is paramount to ensure the appearance of the spectra are as consistent as possible, therefore only spectra collected with the WiFeS setup described above are eligible to be considered standard stars of the classification system. 
The remaining usable spectra come from two other spectrographs (see Table~\ref{tab:setups})
and can be used to assign classes to RCBs, but cannot be reliably used as standards. See Table~\ref{tab:classes} to find which setup was used to observe each star. 

Each spectrum is shifted to the rest frame and dereddened. For the extinction, we searched \citet{Green2019_extinction} for each star to find an A$_V$ value at the distance given by \citet{BailerJones2021_gaiadistances}, and if the star was not present, we used the A$_V$ value from \citet{S_and_F2011_extinction}. The extinction correction was applied using an average R(V) = 3.1 and CCM dust \citep{CCM89_dust}.

Spectroscopic observation of RCB stars in particular is not entirely straightforward, as one must carefully consider the photometric status of the star. Even a small decline ($\sim$0.5 mag) can redden the spectrum and veil the spectral features significantly, with strong declines allowing only the emission spectrum of the surrounding gas
to be observed. Additionally, the phase of an RCB's decline has differing effects on the spectral appearance, a detailed description of which can be found in \citet{Clayton1996_review} and \citet{Rao2004_UWCendecline}. Therefore, we take special care to ensure that all spectra used in classification are taken at or very near to maximum light. The photometric status of each star is listed in Table~\ref{tab:classes}. We include the stars with unknown photometric status currently with an estimated class (indicated by a ``:'' after their class), the caveat being that their classification will change if the star is found to have been in a decline. We additionally include stars with very small (less than 0.5 mag) declines, as this is within the range of an RCB pulsation amplitude. Again, these stars are classified with a  ``:'' after their temperature class to indicate that the classification can only be an estimate until a full maximum light spectrum is observed. Lastly, any incomplete (i.e. not spanning the full WiFeS wavelength coverage) spectra are included in our analysis as estimated classes. Indeed, many of these stars with incomplete spectra are very highly reddened (A$_V$ > 6 mag) and thus their blue spectra are quite difficult to observe. After removing all unusable spectra and two especially peculiar stars (ASAS-RCB-6 and A166, see Section~\ref{sec:conclusions}, para. 5), we are left with 102 total stars with spectra using the full spectral range from WiFeS and 26 stars with either an incomplete spectrum or from a different spectrograph. This leaves us with 128 total HdC stars which can be classified in this work-- 33 dLHdCs and 95 RCBs.

\subsection{New HdC Stars}

In this work we additionally present observations of five new HdC stars that were previously unpublished. V2331 Sgr was previously reported as an S star in \citet{Tisserand2013_asasdiscovery} due to a mislabeling within the data set. V4017 Sgr has been known since 1975 \citep{Hoffleit1975_discovery} however this is the first time its spectrum has been published. WISE J182235.25-033213.2 was considered a strong RCB candidate in \citet{Tisserand2020_plethora}, however we now confirm its status as an RCB  after observing it in a bright phase. Finally, the two stars M38 and P12 are newly discovered Galactic dLHdCs. These were found using the same methods as \citet{Tisserand2022_dlhdcdiscovery} with an increase of the sky search area up to E(B-V)=2.0 mag. We comment that for the star M38, we do find some IR photometric variations in the WISE monitoring survey \citep{WISE}, but the star does not exhibit any clear variability in the optical in ASAS-SN, ATLAS, nor Gaia surveys \citep{ASASSN1,ASASSN2,ATLAS,Gaia_EDR3}. Nevertheless, we consider this star a dLHdC until dust-related optical photometric declines are observed, as was done in \citet{Tisserand2022_dlhdcdiscovery} with similar stars.


\begin{table*}
	\centering
	\caption{Telescope setups \label{tab:setups}}
	\begin{tabular}{lC{3cm}C{3cm}cr} 
		Setup & Telescope & Instrument & Grating Name & Resolving Power\\
		\hline
		W & ANU 2.3m & WiFeS & B3000/R3000 & 3000 \\
		D & ANU 2.3m & Dual Beam Spectrograph & 158R & 500 \\
        T & TNG 3.58m & DOLORES & VHRV/LRR & 1527(blue)/714(red) \\
		\hline
	\end{tabular}
\end{table*}


\section{Methods}
\label{sec:methods}

\subsection{The Precursory MK Process}
\label{subsec:mkprocess}

Traditionally, stellar classification was performed entirely by eye, with automated systems being purposefully avoided. This practice was intended to keep the classification system less vulnerable to our lack of theoretical knowledge and instead rely only on the appearance of the stellar spectra. The practice of creating such a system was coined the ``MK Process'' in \citet{Morgan1984_mkprocess}. This process dictates that the classification system is defined solely by the existence of standard stars within the class, and emphasizes the idea of ``natural groupings'' of the data, i.e., the groupings that the spectra seem to fall into naturally. For HdC stars, the most clear and obvious natural group arises based on the presence or absence of the C$_2$ Swan bands, and in fact these cool and warm groups have been used for many years \citep[e.g.,][]{Tisserand2020_plethora}. We note there is also a hot group of five known RCB stars with T\textsubscript{eff} > 12000 K and we do not include them in this analysis as they have very significant spectral variance between each other \citep{Demarco2002_hotrcbs,Tisserand2020_plethora}. Upon inspection, we found that this first natural group can be further refined by noting that the C$_2$ molecules dominating the blue portion of cooler HdC spectra dissociate at warmer temperatures than the CN molecules in the red portion of the spectra. Thus, the currently used natural groups are the cool (C$_2$ and CN present), mild (only C$_2$ present), and warm (no obvious molecular features) groups, as used in \cite{Tisserand2022_dlhdcdiscovery}.

With these rough natural groupings in mind, we created a smooth spectral sequence using bright stars with published temperature estimates in the literature (listed in Table~\ref{tab:lit_temperatures}). These spectra contained the full spectral range of 3800 to 9000 \text{\AA}. Upon doing so, we noticed that the sequence which ensures the smoothest spectral variance does not correspond to a sequential list based on published temperature estimates. This is not surprising since there are many different temperature estimates found by many different methods, which can vary up to hundreds of degrees K. This is explored further in Section~\ref{subsec:color_calib}. From this exercise, it became clear that the classification system needed two axes: one for temperature and one for the carbon band strength. 

As mentioned previously, automated spectral classification has been historically avoided. However, as large scale spectroscopic surveys such as the Sloan Digital Sky Survey \citep[SDSS;][]{SDSSIV} have become more commonplace, automation has become significantly more important and work has been done on development of classification algorithms. A long-term goal of the authors is to be able to automate classification of HdC stars, differentiating them from traditional carbon stars in these large sky surveys, and we therefore chose to implement the methods of \citet{BailerJones1998_pca} on our HdC dataset to streamline classification which is still finalized by a human classifier.

\begin{table}
	\centering
	\caption{Temperatures found in the literature}
	\label{tab:lit_temperatures}
	\begin{tabular}{lccr} 
		\hline
		Star & T\textsubscript{eff} (K) & Source \\
		\hline
		ES Aql & 5000 & \citet{Yakovina2013} \\ 
		FH Sct & 6250 & \citet{Asplund2000} \\ 
		GU Sgr & 6250 & \citet{Asplund2000}  \\ 
		NSV 11154 & 5200 & \citet{Yakovina2013} \\ 
		R CrB & 6750 & \citet{Asplund2000} \\
		RS Tel & 6750 & \citet{Asplund2000}  \\ 
		RT Nor & 7000 & \citet{Asplund2000} \\ 
		RY Sgr & 7250 & \citet{Asplund2000} \\ 
		RZ Nor & 6750 & \citet{Asplund2000} \\ 
		S Aps & 5400 & \citet{Asplund1997}  \\ 
		SU Tau & 6500 & \citet{Asplund2000} \\ 
		SV Sge & 5000 & \citet{Yakovina2013}  \\ 
		U Aqr & 6000 & \citet{Asplund1997}  \\ 
		UV Cas & 7250 & \citet{Asplund2000} \\ 
		UW Cen & 7500 & \citet{Asplund2000}\\ 
		UX Ant & 7000 & \citet{Asplund2000} \\ 
		V CrA & 6250 & \citet{Asplund2000} \\ 
		V2552 Oph & 6750 & \citet{RaoLambert2003}  \\ 
		V3795 Sgr & 8000 & \citet{Asplund2000} \\ 
		V482 Cyg & 6500 & \citet{Asplund2000}  \\ 
		V854 Cen & 6750 & \citet{Asplund1998_V854Cen}  \\ 
		VZ Sgr & 7000 & \citet{Asplund2000}  \\ 
		WX CrA & 5300 & \citet{Asplund1997}  \\ 
		XX Cam & 7250 & \citet{Asplund2000} \\ 
		Y Mus & 7250 & \citet{Asplund2000}  \\ 
		Z Umi & 5250 & \citet{Kipper2008} \\ 
		HD 137613 & 5400 & \citet{Asplund1997}  \\ 
		HD 148839 & 5625 & \citet{Bergeat2001}  \\ 
		HD 173409 & 6100 & \citet{Asplund1997}  \\ 
		HD 175893 & 5640 & \citet{Bergeat2001}  \\ 
		HD 182040 & 5590 & \citet{Bergeat2001}  \\ 
		\hline
	\end{tabular}
\end{table}

\subsection{Principal Component Analysis}

Automated stellar classification was traditionally difficult due to the high dimensionality of spectral data, however recent work has begun to address this problem. Following the methods outlined by \citet{BailerJones1998_pca}, we first reduce the dimensionality of our data set. Specifically, we perform a principle component analysis (PCA) on the spectral data. PCA is a method of calculating the basis along which the most variance in the data can be explained, and transforming the data into a linear combination of the new basis spectra, or eigenspectra. As the principal components (the eigenspectra) are arranged such that the first of these components explains the largest source of variance in the data, many of the later components can be ignored without significant reduction of the conveyed information. Thus, the dimensionality of each spectrum can be reduced from the order of thousands to the order of tens.

\begin{figure*}
	\includegraphics[width=\textwidth]{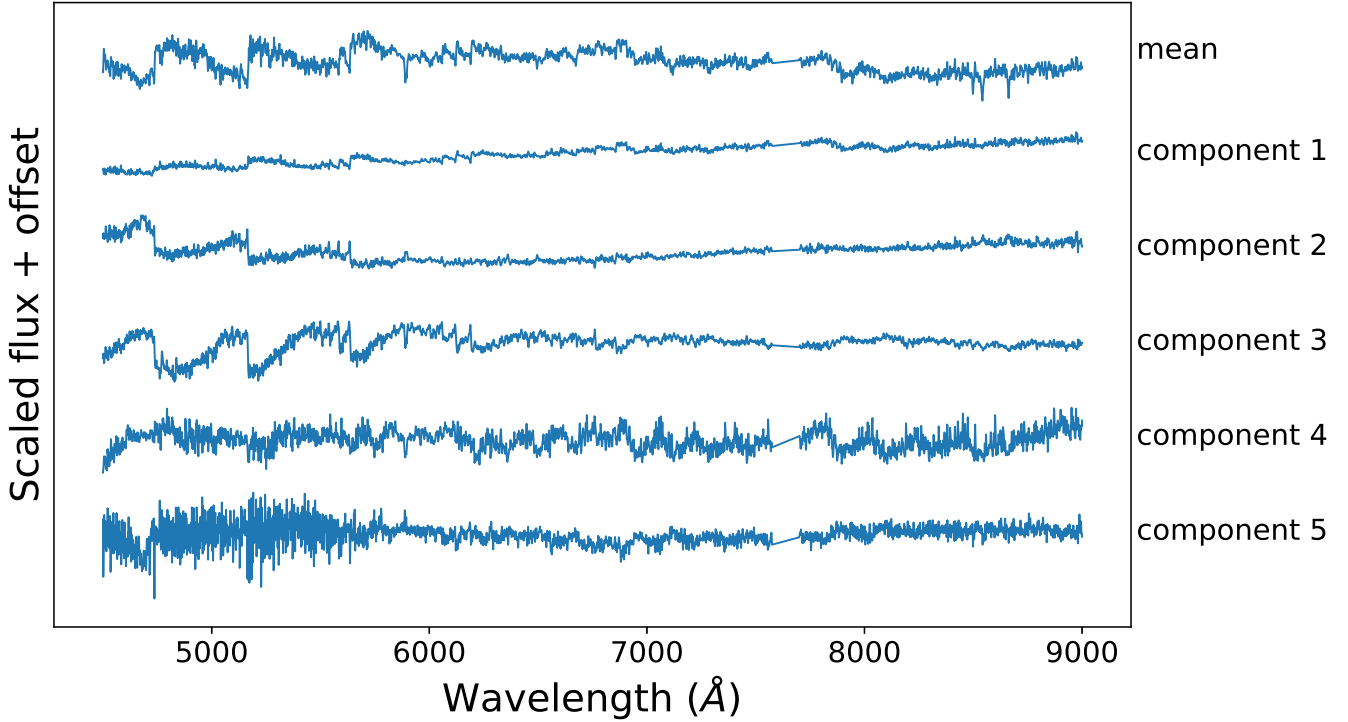}
    \caption{The mean HdC spectrum from 4500 to 9000 \text{\AA} is shown at the top with the first 5 principal components offset below it. The component spectra, or eigenspectra, are labeled such that component 1 explains the most spectral variance, and each subsequent eigenspectrum explains less variance than the eigenspectrum before it. The spectral region from 7575 to 7700 \text{\AA} is removed to avoid the telluric B band.}
    \label{fig:5components}
\end{figure*}

We performed a PCA on all maximum light HdC spectra that have data from 4500 to 9000 \text{\AA} using the \texttt{PCA} algorithm from {\sc scikit-learn} \citep{scikit-learn}. Note that this spectral range does not extend as far into the blue as the maximum WiFeS spectral range, as dereddening the spectra magnifies the noise in the bluest part of the spectrum, and the PCA algorithm places too much emphasis on trying to explain this noise rather than the important spectral features. We used the first 50 principal components, as by that point we are explaining 99.6\% of the cumulative variance in the data set.
The first five principal components and the mean spectrum are shown in Figure~\ref{fig:5components}. We can see that important spectral features are spread across multiple eigenspectra (e.g., the C$_2$ bands can be seen clearly in the first three components) and that each eigenspectrum contains multiple types of features (e.g., C$_2$ and CN bands can both be seen in the same eigenspectrum). However, we can still attribute some components most strongly to certain spectral properties. Component 1 most strongly correlates to the shape of the underlying continuum and components 2 and 3 most strongly denote the strength of the C$_2$ and CN bands. Thus, we confirm our intuition that the continuum and the carbon molecular bands are the most important features upon which to classify the HdC spectra.

Once the data have all been deconstructed into their principal components, the dataset forms a 50 dimensional surface, for which the first three dimensions are visualized in Figure~\ref{fig:kmeans_groups}. As suspected, components 1 and 2 are strongly correlated, as they correspond most closely to the shape of the continuum and the presence of molecular bands, respectively. This effect is mostly due to the differences in temperature in the stars, which will both change the shape of the continuum and affect the strength of the molecular band features. There is more spread in component 3, and it correlates more with component 1 than component 2. We find that component 3 most closely represents the strength of the C$_2$ bands, although there is a component of the C$_2$ band strength within the other two components as well. From Figure~\ref{fig:5components}, we can see that the C$_2$ bands are represented in component 1, and the inverse C$_2$ bands are seen in components 2 and 3. Therefore, we expect the C$_2$ band strength to be best represented by a combination of the first three components.

\subsection{K-means Clustering}

\begin{figure}
	\includegraphics[width=\columnwidth]{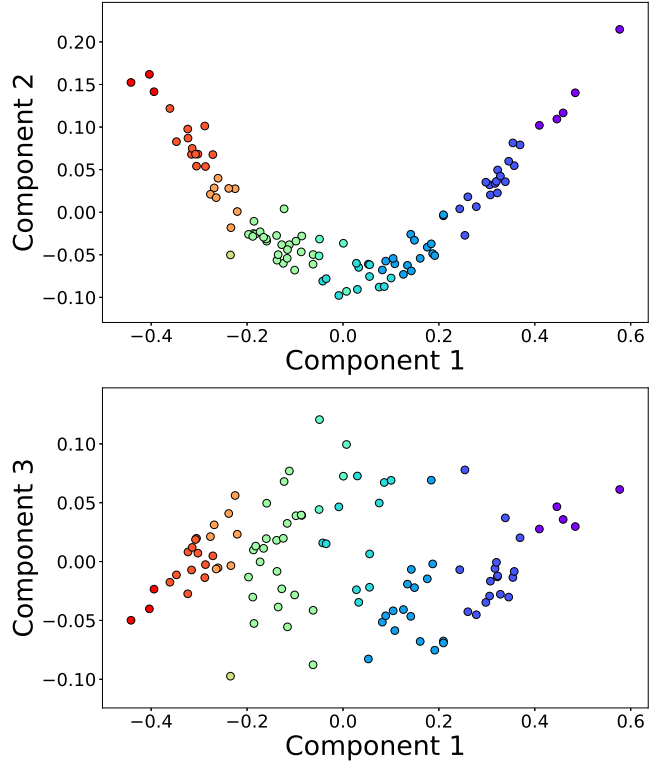}
    \caption{The first two projections of the PCA data set overlaid with the K-means clusters. The colors of the points correspond to the 10 different initial K-means clusters, where the purple indicates cooler stars and the red indicates warmer stars.}
    \label{fig:kmeans_groups}
\end{figure}

Once the classification set had undergone dimensionality reduction, we performed a K-means clustering on the PCA deconstruction of our dataset. K-means is a simple, iterative clustering algorithm which defines cluster centers and groups the data based on which cluster center is closest to the datum, readjusting the position of the cluster centers upon each iteration. We initialized the cluster centers using the locations of ten bright, well-known stars over a range of temperatures. Those stars were RZ Nor, V517 Oph, S Aps, B563, WISE J194218.38-203247.5, SV Sge, HD 175893, RS Tel, C528, RY Sgr, Y Mus, and HD 182040. The results of this clustering can be seen in Figure~\ref{fig:kmeans_groups}. 

One can see that the K-means clustering captures the differences in temperature smoothly across the set of classification spectra. However, K-means assumes evenly spatially sized, spherical clusters, which does not match the shape in our data set. One can also see that the clusters created have nearly excluded stars that lie far from the mean in the component 3 axis, placing them into their own classes. While other clustering algorithms may be more adept at handling non-uniform distributions of data, these algorithms struggle with the high dimensionality of our set. Therefore, we feel the K-means algorithm suits our science well, as it provides a simple baseline for which a human classifier can easily make adjustments to. From the results, it is clear that K-means is nonetheless effective at gathering our stars into groups of similar temperature. We note that the K-means clustering is not robust to a random initialization of cluster centers. A well-informed set of initial cluster centers is necessary for the K-means algorithm to provide a useful springboard to a proper classification system. Human intervention is still required to ensure such a system is useful even after a clustering algorithm is applied.

\subsection{Final Classification}
\label{subsec:final}

\begin{figure}
	\includegraphics[width=\columnwidth]{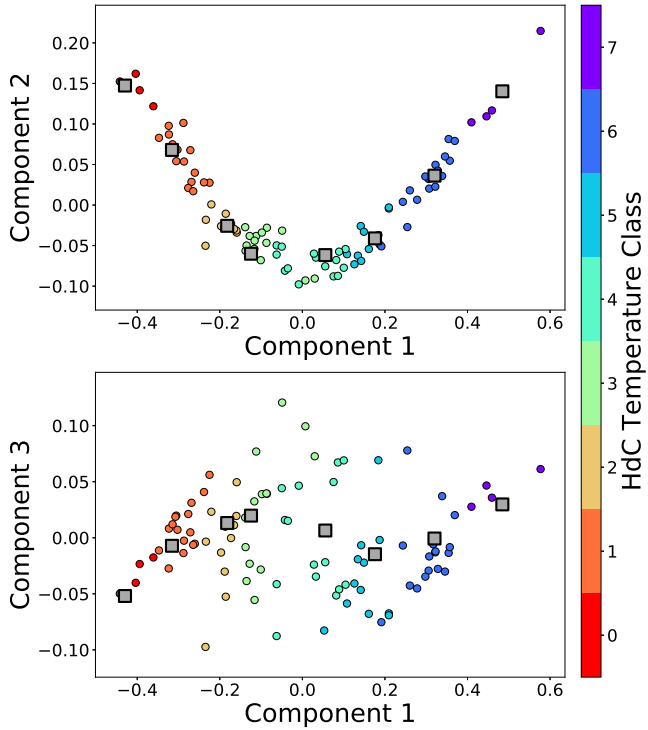}
    \caption{The first two projections of the PCA data set overlaid with the final classification of stars. The colors of the points correspond to the eight different temperature classes, where the purple indicates cooler stars and the red indicates warmer stars. The grey squares indicate the location of the standard star used for each of the classes, with HdC0 being the leftmost standard and HdC7 being the rightmost standard.}
    \label{fig:final_projection}
\end{figure}

\begin{figure}
	\includegraphics[width=\columnwidth]{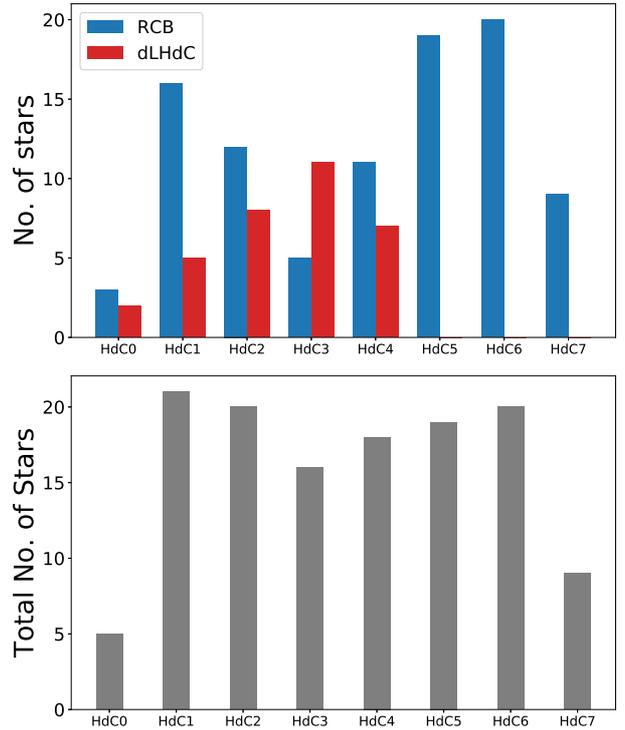}
    \caption{In the upper panel is a histogram showing the number of dLHdCs (red) and RCBs (blue) in each class, and in the lower panel is a histogram showing the total size of each class.}
    \label{fig:histogram}
\end{figure}

\begin{figure*}
	\includegraphics[width=\textwidth]{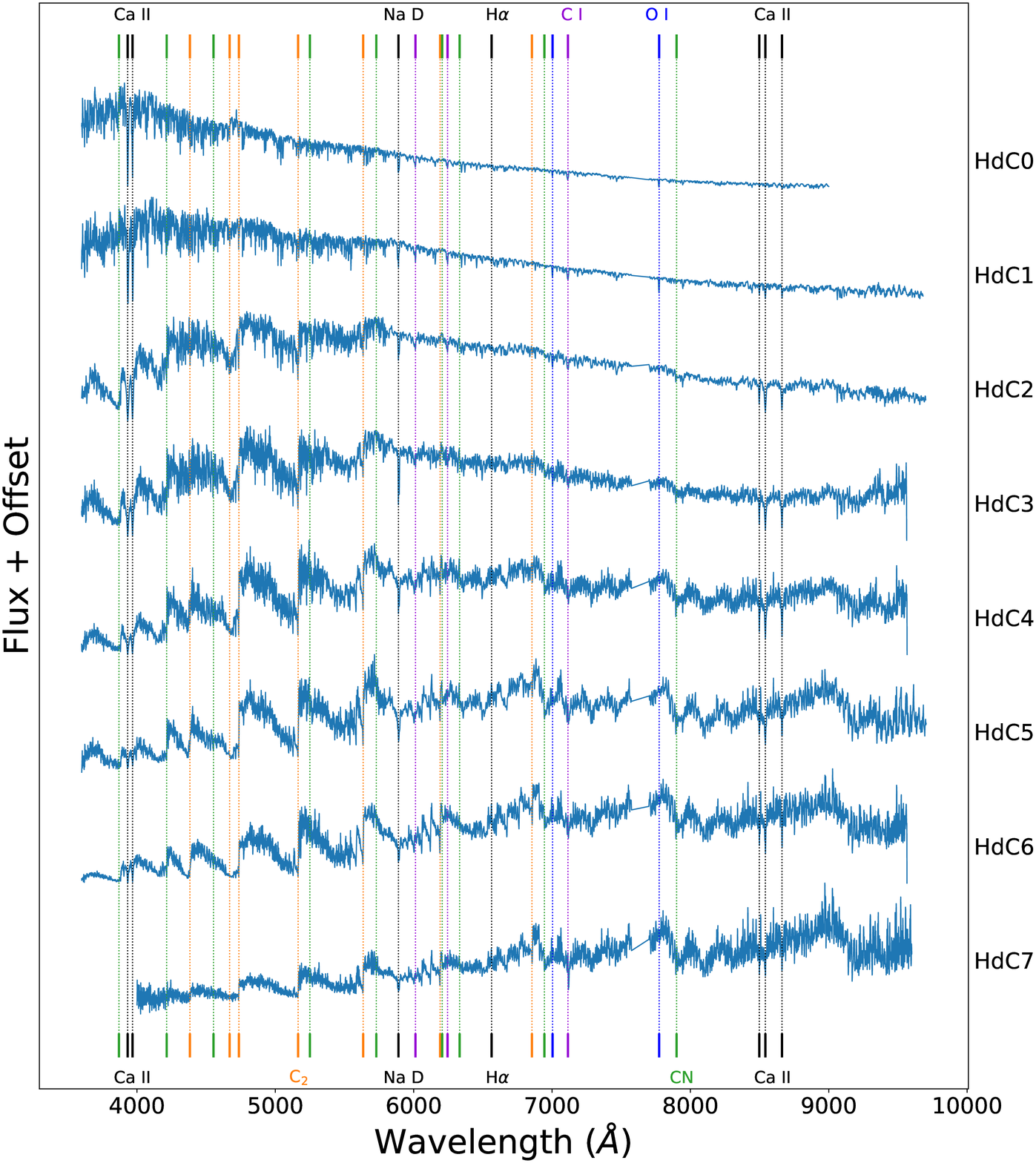}
    \caption{The spectra of the standard stars for each of the temperature classes. The HdC0 class at the top indicates the warmest group of stars, and HdC7 at the bottom indicates the coolest group of stars. Each spectrum is divided by its maximum flux value and offset from each other. The region around the telluric B band (7575 to 7700 \text{\AA}) is removed. We additionally label a few of the most important spectral features, namely Ca H \& K, the Ca II IR triplet, the rarely seen H$\alpha$, and Na D in black, the strongest C I features in purple, the strongest O I features in blue, the C$_2$ bandheads in orange, and the CN bandheads in green. The features labeled at the top indicate those most easily observed in the warm stars, and those labeled at the bottom are features best seen in cool stars.}
    \label{fig:sequence}
\end{figure*}

The implementation of PCA and K-means clustering provided the framework for a new classification system. The final classification system was created by adjusting the classes created by clustering until the system varied smoothly. In order to preserve the intent of the MK process, this process was done without any information on the physical processes driving the spectral differences. However, it is clear that the parameter which affects the spectra most strongly is the temperature, and therefore the classification system's main axis is a temperature axis. We thus ended up with a total of eight classes that form a continuum in temperature space. Their projections in the principal component basis can be seen in Figure~\ref{fig:final_projection}. The spectral continuum itself can be seen in Figure~\ref{fig:sequence}. The number of classes was reduced from 10 in the K-means clusters to 8 in the final classification system as we found that two of the K-means clusters lay outside the distribution of most of the HdC stars, and eight classes were sufficient to fully sample the distribution of temperatures, and thus spectral appearances, at this resolution. In Figure~\ref{fig:histogram}, one can see that there is a nearly flat distribution of stars in each class, which is not currently straightforward to explain.

\begin{figure*}
	\includegraphics[width=\textwidth]{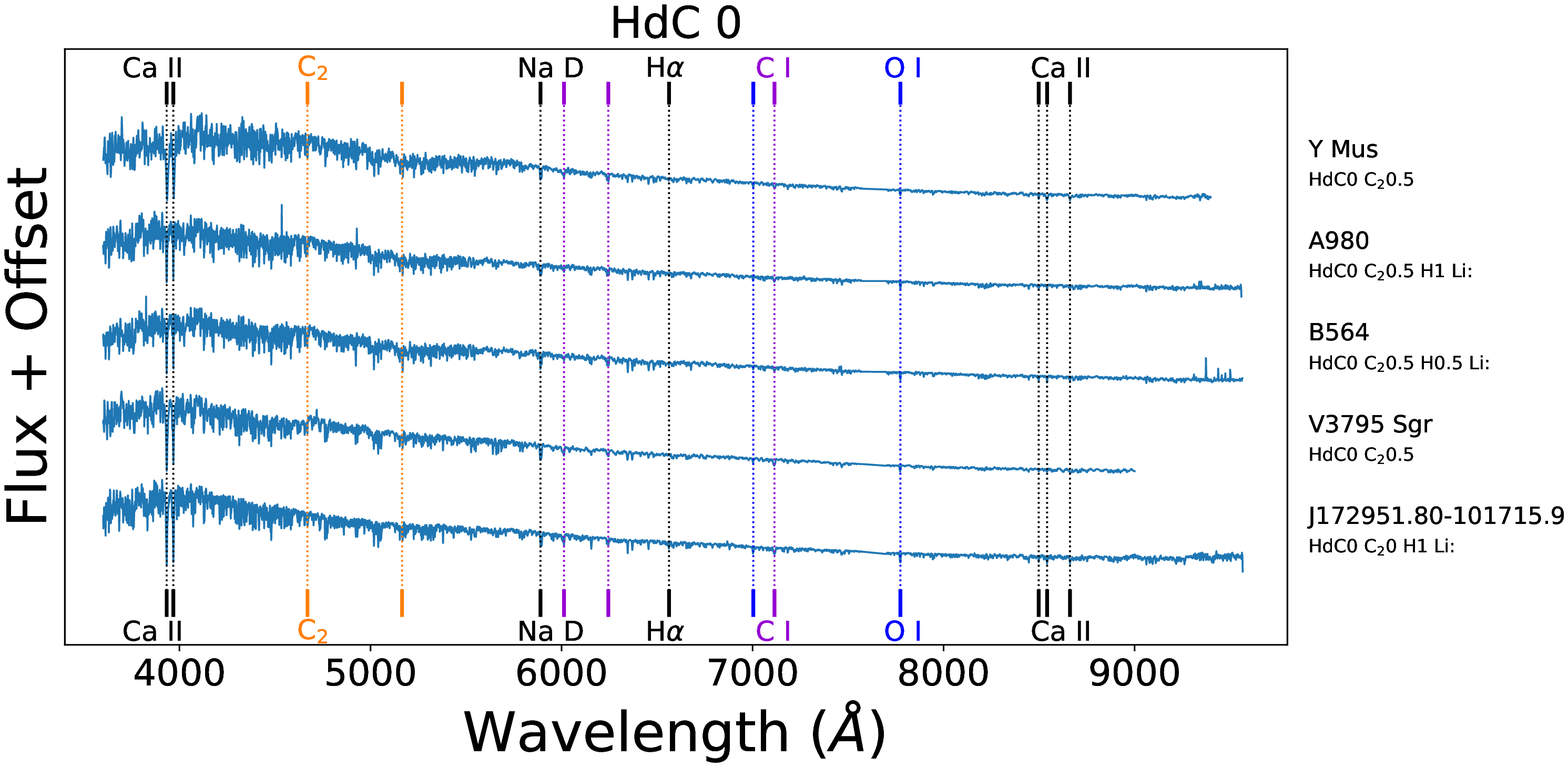}
    \caption{The spectra of the HdC0 class of stars. The stars are labeled on the right hand side with their class listed below. The spectra are arranged such that the star at the top has the strongest C$_2$ bands and the star at the bottom has the weakest. Each spectrum is divided by its maximum flux value and offset from each other. The spectral region from 7575 to 7700 \text{\AA} is removed to avoid the telluric B band. We additionally label a few of the most important spectral features for HdC0, namely Ca H \& K, the Ca II IR triplet, the rarely seen H$\alpha$, and Na D in black, the strongest C I features in purple, the strongest O I features in blue, and the C$_2$ bandheads visible at this temperature in orange. There are no visible CN bands at this temperature.}
    \label{fig:set0}
\end{figure*}

\begin{figure*}
	\includegraphics[scale=0.5]{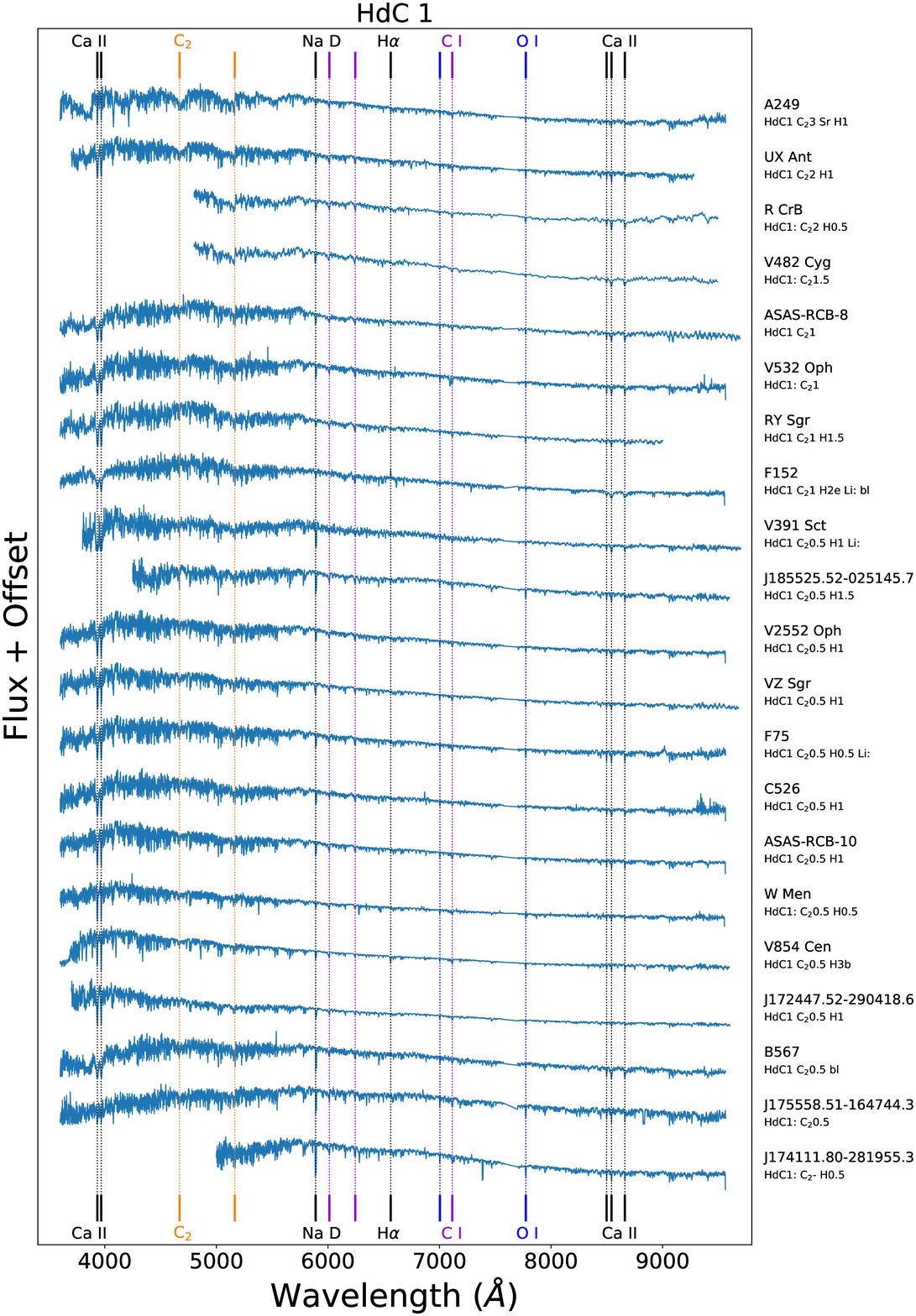}
    \caption{The spectra of the HdC1 class of stars. The information is arranged in the same way as Figure~\ref{fig:set0}. We label a few of the most important spectral features for HdC1, namely Ca H \& K, the Ca II IR triplet, the rarely seen H$\alpha$, and Na D in black, the strongest C I features in purple, the strongest O I features in blue, and the C$_2$ bandheads visible at this temperature in orange. There are no visible CN bands at this temperature.}
    \label{fig:set1}
\end{figure*}

\begin{figure*}
	\includegraphics[scale=0.5]{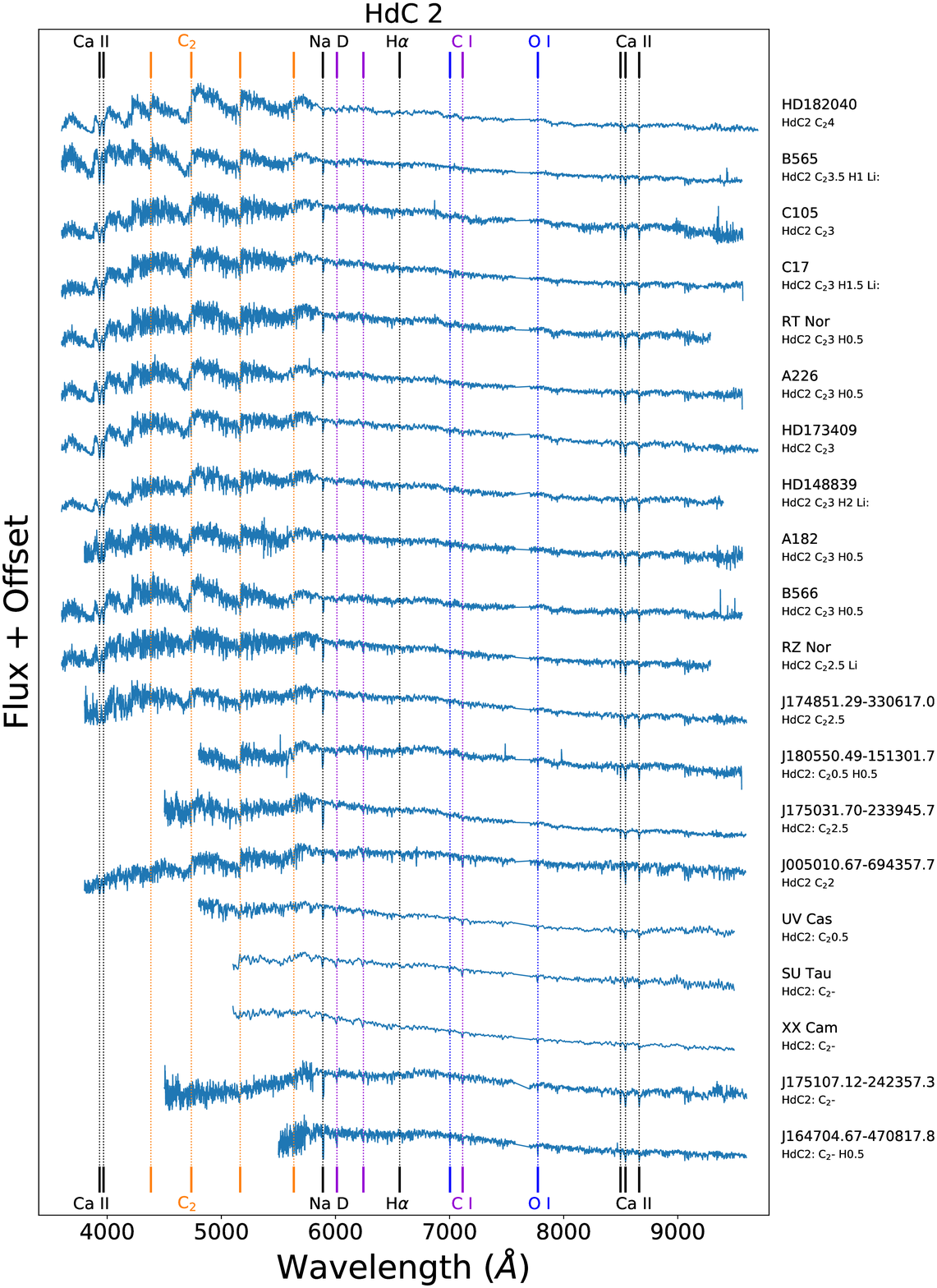}
    \caption{The spectra of the HdC2 class of stars. The information is arranged in the same way as Figure~\ref{fig:set0}. We additionally label a few of the most important spectral features for HdC2, namely Ca H \& K, the Ca II IR triplet, the rarely seen H$\alpha$, and Na D in black, the strongest C I features in purple, the strongest O I features in blue, and the C$_2$ bandheads visible at this temperature in orange. There are no visible CN bands at this temperature.}
    \label{fig:set2}
\end{figure*}

\begin{figure*}
	\includegraphics[width=\textwidth]{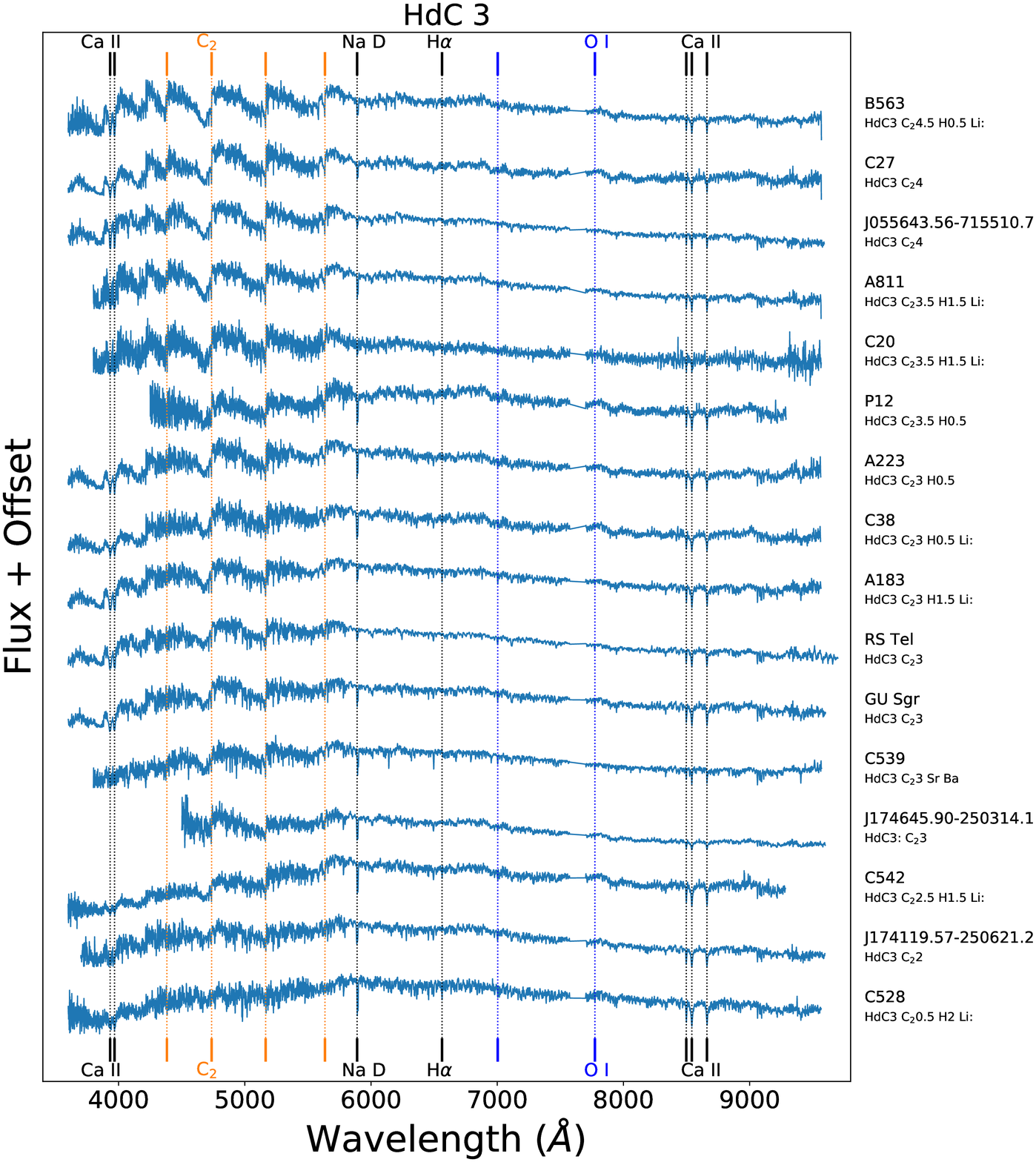}
    \caption{The spectra of the HdC3 class of stars. The information is arranged in the same way as Figure~\ref{fig:set0}. We additionally label a few of the most important spectral features for HdC3, namely Ca H \& K, the Ca II IR triplet, the rarely seen H$\alpha$, and Na D in black, the strongest O I features in blue, and the C$_2$ bandheads visible at this temperature in orange. There are no visible CN bands nor strong C I features at this temperature.}
    \label{fig:set3}
\end{figure*}

\begin{figure*}
	\includegraphics[scale=0.525]{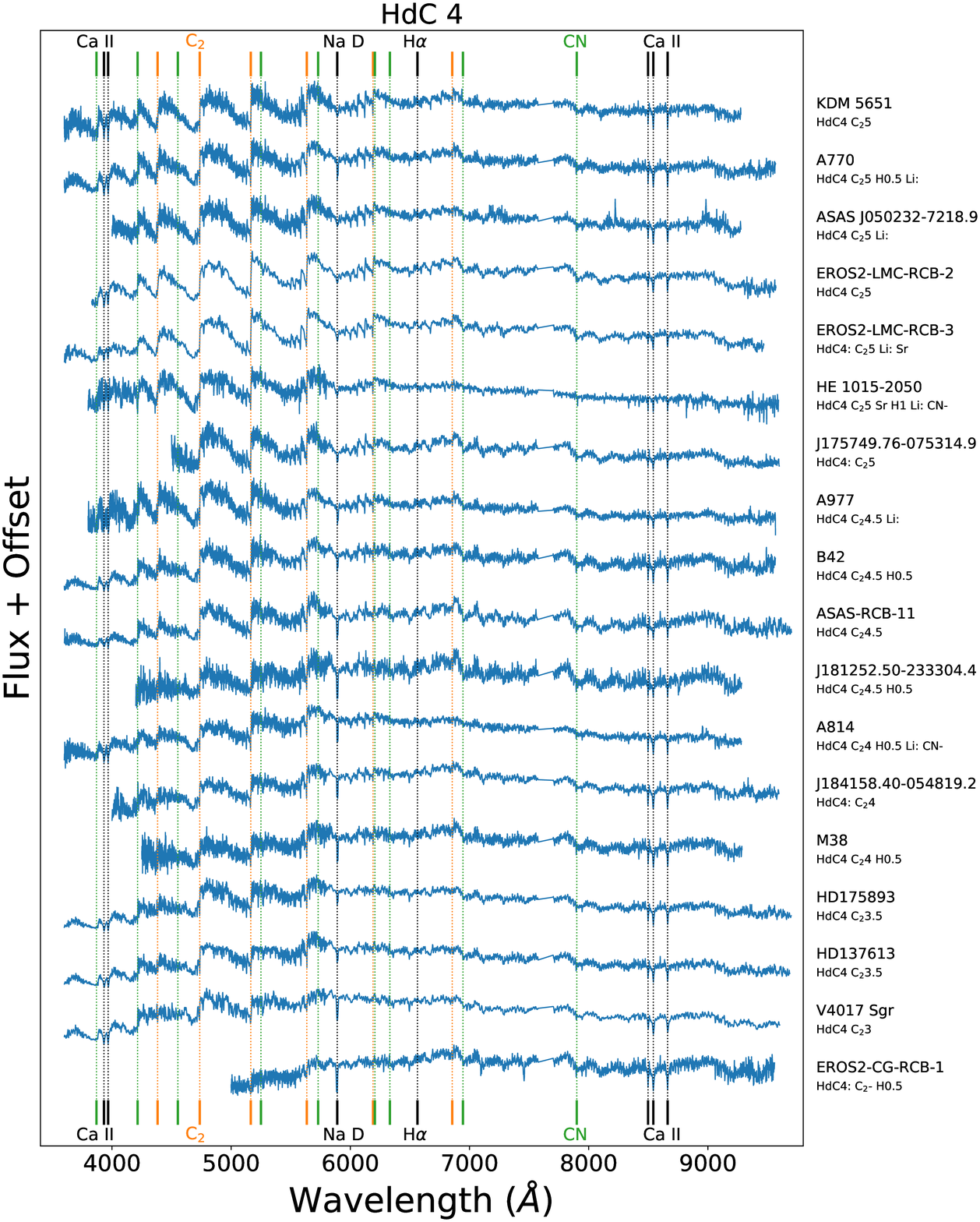}
    \caption{The spectra of the HdC4 class of stars. The information is arranged in the same way as Figure~\ref{fig:set0}. We additionally label a few of the most important spectral features for HdC4, namely Ca H \& K, the Ca II IR triplet, the rarely seen H$\alpha$, and Na D in black, the C$_2$ bandheads in orange, and the CN bandheads in green. The neutral atomic features are no longer easily visible at this temperature.}
    \label{fig:set4}
\end{figure*}

\begin{figure*}
	\includegraphics[scale=0.5]{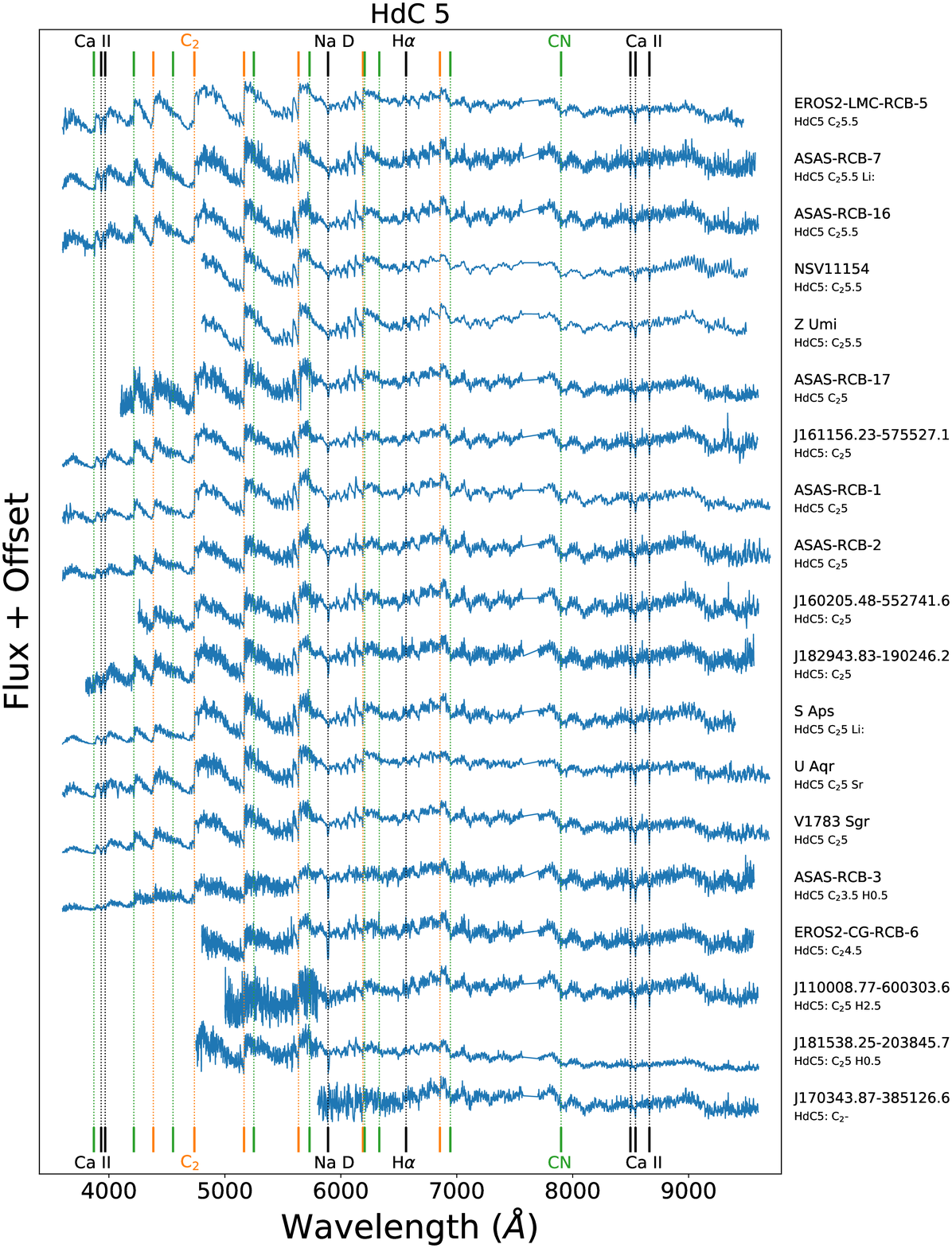}
    \caption{The spectra of the HdC5 class of stars. The information is arranged in the same way as Figure~\ref{fig:set0}. We additionally label a few of the most important spectral features for HdC5, namely Ca H \& K, the Ca II IR triplet, the rarely seen H$\alpha$, and Na D in black, the C$_2$ bandheads in orange, and the CN bandheads in green. The neutral atomic features are no longer easily visible at this temperature.}
    \label{fig:set5}
\end{figure*}

\begin{figure*}
	\includegraphics[scale=0.525]{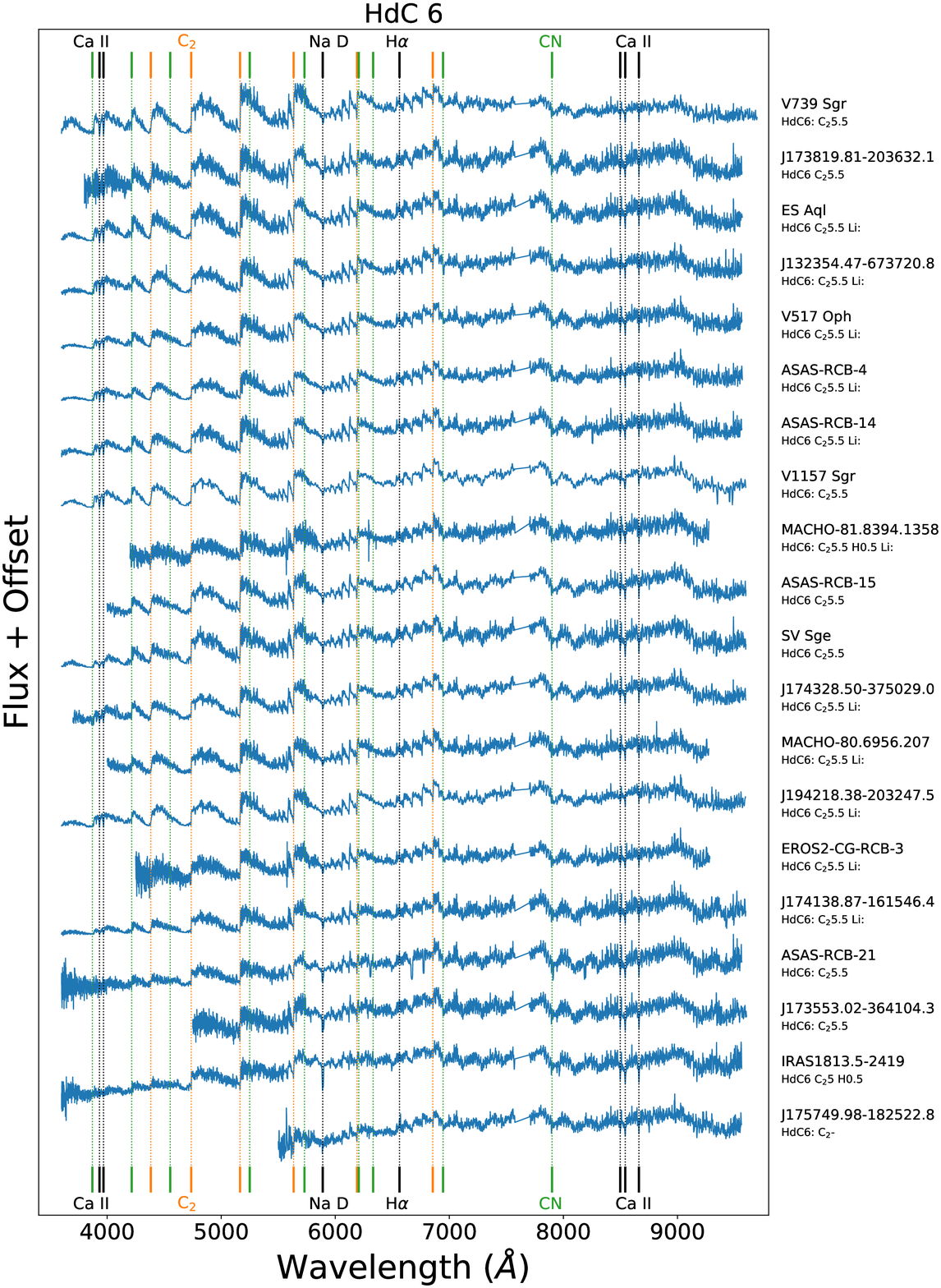}
    \caption{The spectra of the HdC6 class of stars. The information is arranged in the same way as Figure~\ref{fig:set0}. We additionally label a few of the most important spectral features for HdC6, namely Ca H \& K, the Ca II IR triplet, the rarely seen H$\alpha$, and Na D in black, the C$_2$ bandheads in orange, and the CN bandheads in green. The neutral atomic features are no longer easily visible at this temperature.}
    \label{fig:set6}
\end{figure*}

\begin{figure*}
	\includegraphics[width=\textwidth]{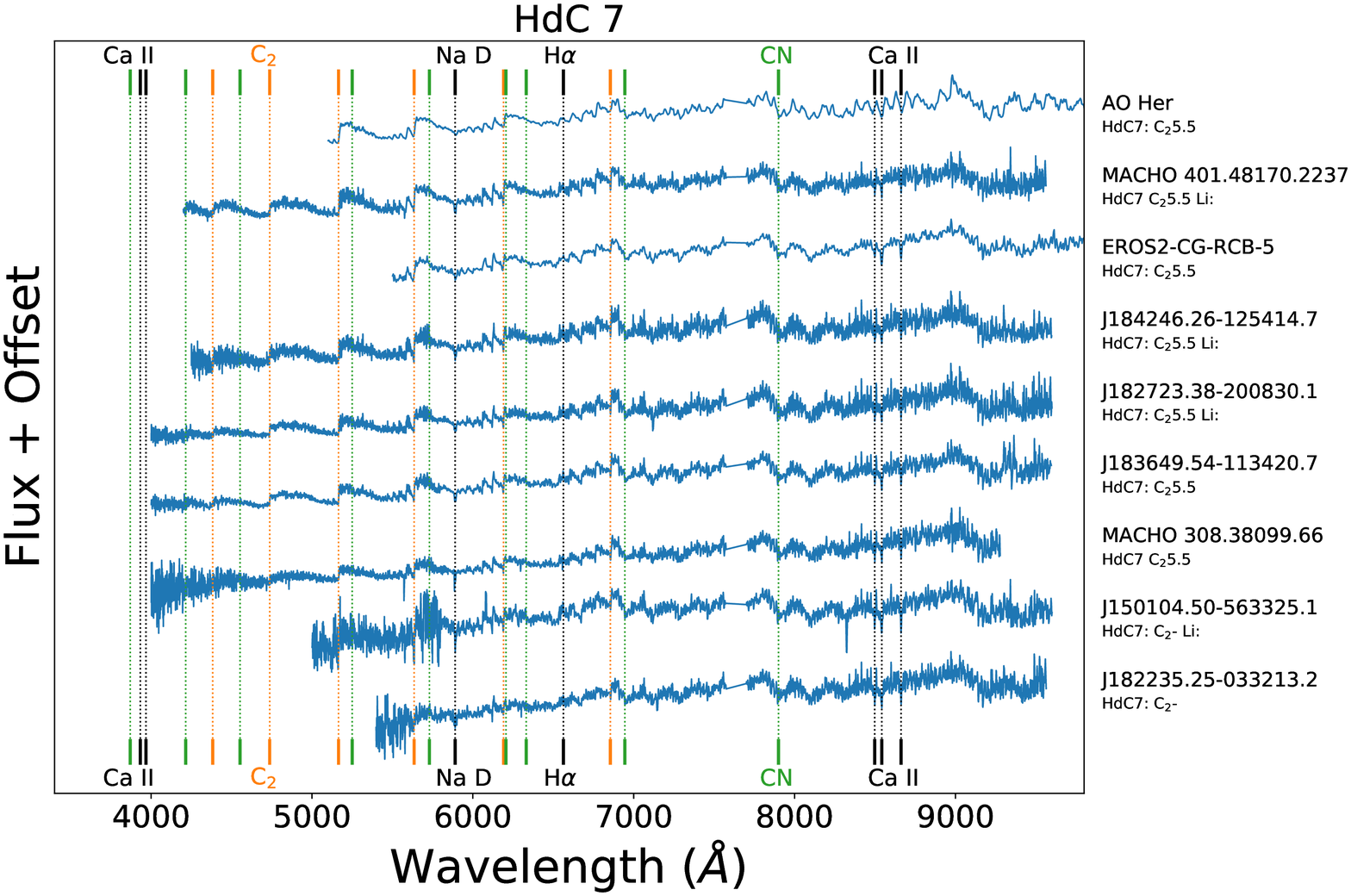}
    \caption{The spectra of the HdC7 class of stars. The information is arranged in the same way as Figure~\ref{fig:set0}. We additionally label a few of the most important spectral features for HdC7, namely Ca H \& K, the Ca II IR triplet, the rarely seen H$\alpha$, and Na D in black, the C$_2$ bandheads in orange, and the CN bandheads in green. The neutral atomic features are no longer easily visible at this temperature.}
    \label{fig:set7}
\end{figure*}

Additionally, we noticed that in each temperature class there is a range of C$_2$ band strengths, especially regarding the disappearance of the 4383 \text{\AA} band in stars with weaker C$_2$ strengths. Therefore, we include an index to indicate the strength of the C$_2$ Swan bands, as in the \citet{Keenan1993_carbonclass} system for carbon stars. This index is a number which ranges from 0 to 6, indicating the Carbon band strengths, and would be denoted via ``C$_2$\emph{y}'' where \emph{y} is the carbon index. Thus, the full class for a star would indicate a temperature class and a carbon index, i.e. ``HdC\emph{x} C$_2$\emph{y}''. In order to aid classification for those accustomed to other systems, we used the stars from the \citet{Barnbaum1996_carbon} carbon star atlas to calibrate our C$_2$ indices. The carbon abundance sequences can be seen in Figures~\ref{fig:set0}-\ref{fig:set7}. 

Standard stars for each class were chosen with preference to bright stars with complete spectra and those near the center of each class in the PCA projection. We also placed a slight preference on dLHdCs, since they are less variable than the RCBs. The standard stars from warmest to coolest are V3795 Sgr, VZ Sgr, HD 173409, A223, B42, ASAS-RCB-2, V517 Oph, and WISE J182723.38-200830.1. We note that the HdC7 standard, WISE J182723.38-200830.1 is in an unknown photometric phase. It is possible that the standard star will change upon reobservation if necessary. One may notice that the standard stars, when shown in Figure~\ref{fig:final_projection}, follow a nearly sinusoidal distribution in the principal component basis. This was not an intentional choice on the authors. Rather, we display a preference to the stars nearest to 0.00 in the component 3 axis. The overall distribution of stars in the lower panel of this figure could either be interpreted as a distribution with a large vertical spread near the center or as two nearly-linear, disjointed sections. We do not distinguish between either interpretation, as we believe the lack of stars around 0.0 in the component 1 and 3 axes is simply due to an overall lack of stars in the sample.

For the initial classification, we used only spectra taken with the WiFeS spectrograph, and had the full spectral range of 4500 - 9000 \text{\AA}. However, we additionally have 27 spectra from other spectrographs with different resolutions or with different wavelength coverage, and we estimate their classes as well. Stars with different resolutions can be classified via their PCA projection, so long as they have the full spectral range, however those with only partial spectra must be classified by direct comparison to the HdC standards. We include these spectra with the others in their respective classes, however it must be noted that we cannot comment on spectral peculiarities outside of the star's spectral range, and therefore the additional peculiarities are subject to appear as we obtain complete spectra. Therefore, we place a ``:'' after their temperature class to indicate that the classification can only be an estimate until a full maximum light spectrum can be viewed. In addition, many of these partial spectra do not contain sufficient information to assign a carbon index, and are therefore denoted by ``C$_2$-''.

It is worth noting that the spectral range for this system is indeed larger than a typical classification region. We chose such a region so as to include the C$_2$ Swan bands in the blue and the atomic features prevalent in the red portion of the spectrum. If one were to perform the above analysis on simply the red portion of the spectrum, the main axis of classification (surface temperature) does not change, however much of the information on the strength of the carbon bands is lost. As an additional test, we performed a similar PCA decomposition on only the 5800 to 7000 \text{\AA} range of the spectra (see Figure~\ref{fig:sequence_zoom}). As expected, we again found that the representation of stellar temperature was unchanged, but lost significant information on the carbon molecular strength. However, a classification scheme based only on the blue portion of the spectrum is also problematic as many of the coolest HdCs (Figures~\ref{fig:set6} and~\ref{fig:set7}) have very little flux in the blue, especially when considering that these stars can be highly reddened if observed in the plane of the galaxy. This is indeed why many of our stars are missing the blue-most portions of their spectra, and why many cool star spectra appear quite noisy blueward of $\sim$4000 \text{\AA}.


\onecolumn
\begin{landscape}
\begin{longtable}{lccccllll}
	\caption[]{HdC Spectral Classes at or near maximum light.} \\ %
	\label{tab:classes}
    Name      & RA & Dec & Instru- &  A$_V$ & Temp. & Class & Photometric & References     \\ %
          & (J2000) & (J2000) & -ment &  (mag)  &  group &  &  state (mag) &  \\ %
         & &  &   &     &   &  &  + epoch {\tiny (D/M/Y) }&  \\ %
\hline
\endfirsthead
\caption[]{(continued)}\\
    Name     & RA & Dec & Instru- &  A$_V$ & Temp. & Class  & Photometric & References     \\ %
          & (J2000) & (J2000) & -ment &  (mag)  & group &  &  state (mag) &  \\ %
          & & &   &     &   &  &  + epoch  {\tiny (D/M/Y)} &  \\ %
\hline
\endhead

		  \multicolumn{9}{c}{} \\ %
		  \multicolumn{9}{c}{Galactic RCB stars} \\ %
         \hline
        AO Her	&	17:35:36.28	&	50:24:39.86	&	T &	0.07	&  Cold  &	HdC7: C$_2$5.5		& Max. {\tiny (20/11/20)}& \small{W24, O14}   \\ 
        ASAS-RCB-1	&	15:44:25.07	&	-50:45:01.26	&	W & 3.61 &  Cold   &	HdC5	C$_2$5		& Max. {\tiny (15/07/10)}  & \small{T13}   \\ %
        ASAS-RCB-2$^{Std}$	&	16:41:24.75	&	-51:47:43.43	& W & 2.12 &  Cold     &	HdC5	C$_2$5		& Max. {\tiny (15/07/10)}  & {\small T13}    \\ 
        ASAS-RCB-3	&	16:54:43.61	&	-49:25:55.43	& W & 2.38 &  Cold     &	HdC5 C$_2$3.5 H0.5	& Max. {\tiny (22/07/22)}  & {\small M12, T13}    \\ 
        ASAS-RCB-4	&	17:05:41.25	&	-26:50:03.36	& W & 0.45 &  Cold     &	HdC6 C$_2$5.5 Li: & Max. {\tiny (22/07/22)}  & {\small M12, T13}		\\ 
        ASAS-RCB-5	&	17:52:25.51	&	-34:11:28.16	&  & 2.16 &  Cold     &	&  & {\small T13}     \\ 
        ASAS-RCB-6$^{Pec}$	&	20:30:04.97	&	-62:07:59.22	&  & 0.13 &  Cold     &				&   & {\small M12, T13}	\\ %
        ASAS-RCB-7	&	17:49:15.66	&	-39:13:16.58	& W & 1.13 &  Cold     &	HdC5 C$_2$5.5 Li:		& Max. {\tiny (22/07/22)}  & {\small T13}     \\ 
        ASAS-RCB-8	&	19:06:39.87	&	-16:23:59.15	& W & 0.41 & Warm    &	HdC1	C$_2$1		& Max. {\tiny (15/07/10)}  & {\small T13}     \\ %
        ASAS-RCB-9	&	16:22:28.83	&	-48:35:55.68	&  & $>$6  &  Cold     &		&  & {\small T13}     \\ 
        ASAS-RCB-10	&	17:01:01.41	&	-50:15:34.90	& W  &	1.78	& Warm    &	HdC1 C$_2$0.5 H1			& Max.  {\tiny (22/07/22)} & {\small T13}     \\ 
        ASAS-RCB-11	&	18:12:03.69	&	-28:08:36.23	& W & 1.64 &  Cold     &	HdC4	C$_2$4.5		& Max. {\tiny (15/07/10)} & {\small T13}     \\ %
        ASAS-RCB-12	&	17:01:01.41	&	-50:15:34.90	&  & 1.53 & Mild    &	&  & {\small T13}     \\ 
        ASAS-RCB-13	&	18:24:43.47	&	-45 24 43.75	&   &	0.18	&  Cold     &				&  & {\small T13}     \\ %
        ASAS-RCB-14	&	16:47:29.74	&	-15:25:22.9	& W & 1.00 &  Cold     &	HdC6 C$_2$5.5 Li:		& Max. {\tiny (15/07/22)} & {\small T13}     \\ 
        ASAS-RCB-15	&	17:08:27.19	&	-32:26:49.60	& W & 1.57 &  Cold     &	HdC6	C$_2$5.5		& Max. {\tiny (06/06/12)} & {\small T13}     \\ %
        ASAS-RCB-16	&	17:14:14.46	&	-21:26:13.78	& W & 1.90 &  Cold     &	HdC5	C$_2$5.5		& Max. {\tiny (06/06/12)}  & {\small T13}     \\ %
        ASAS-RCB-17	&	17:17:44.49	&	-29:38:00.20	& W & 3.16 &  Cold     &	HdC5	C$_2$5		& Max. {\tiny (06/06/12)}  & {\small T13}     \\ %
        ASAS-RCB-18 &	19:00:09.44	&	-02:02:57.90    &  & 3.79 &  Cold   &   &   & {\small T13}     \\ 
        ASAS-RCB-19	&	19:01:33.68	&	14:56:09.59	&  & 2.23 &  Cold     &			&  & {\small T13}     \\ 
        ASAS-RCB-20	&	19:53:43.15	&	14:41:09.36	&  & 0.71 &  Cold     &		&  & {\small T13}     \\ 
        ASAS-RCB-21	&	18:58:41.80	&	-02:20:11.30	&  W &	3.87 &  Cold     &	HdC6: C$_2$5.5 & $\Delta\sim$0.5 {\tiny (15/07/22)}  & {\small T13}     \\ 
        C105	&	19:39:35.48	&	34:34:47.19	& W & 0.74 &  Mild  &	HdC2	C$_2$3		& Max. {\tiny (29/06/21)}  & {\small T22}   \\ %
        EROS2-CG-RCB-1	&	17:52:20.04	&	-29:03:31.69	&  W &	6.25	&  Mild  &		HdC4: C$_2$- H0.5 & Max. {\tiny (15/07/22)} & {\small T08}   \\ 
        EROS2-CG-RCB-3	&	17:58:28.30	&	-30:51:16.63	& W & 3.40  &  Cold    &	HdC6	C$_2$5.5	Li:	& Max.  {\tiny (16/03/22)}  & {\small T08}    \\ %
        EROS2-CG-RCB-4	&	17:46:16.23	&	-32:57:40.9	&    &  5.35	&  Cold    &		&  & {\small T08}    \\ 
        EROS2-CG-RCB-5	&	17:46:00.34	&	-33:47:56.69	&   D &  3.65	&  Cold    &	HdC7: C$_2$5.5		& Max. {\tiny (18/04/07)}  & {\small T08}    \\ 
        EROS2-CG-RCB-6	&	17:30:23.85	&	-30:08:28.50	&   W &  6.52	&  Cold    &	HdC5: C$_2$4.5	& Max. {\tiny (22/07/22)}  & {\small T08}    \\ 
        EROS2-CG-RCB-7	&	17:29:37.11	&	-30:39:36.85	&    & $>$6  &  Cold    &		&  & {\small T08}    \\ 
        EROS2-CG-RCB-8	&	17:39:20.73	&	-27:57:22.68	&   &	5.87	&  Cold    &				&  & {\small T08}    \\ %
        EROS2-CG-RCB-9	&	17:48:30.90	&	-24:22:56.7	&    &  $>$6	&  Cold    &		&  & {\small T08}    \\ 
        EROS2-CG-RCB-10	&	17:45:31.41	&	-23:32:24.4	&   &	3.31	&  Cold    &				&  & {\small T08}    \\ %
        EROS2-CG-RCB-11	&	17:48:41.57	&	-23:00:26.7	&    &  3.46 &  Cold    &		&  & {\small T08}    \\ 
        EROS2-CG-RCB-12	&	17:19:58.51	&	-30:04:21.51	&   &	4.28	&  Cold    &				&  & {\small T08}    \\ %
        EROS2-CG-RCB-13	&	18:01:58.25	&	-27:36:48.40	&  & 3.35  &  Cold    &  &  & {\small T08}    \\ 
        EROS2-CG-RCB-14	&	18:13:14.87	&	-27:49:41.04	&    &  1.56	&  Cold    &	&  & {\small T08}    \\ 
        ES Aql	&	19:32:21.61	&	-00:11:30.96	& W & 1.08  &  Cold    &  HdC6 C$_2$5.5 Li: & Max.	{\tiny (16/07/22)}  & {\small C02}	\\ 
        FH Sct	&	18:45:14.85	&	-09:25:36.11	&  & 2.49  & Warm    &			&  & {\small T81, F92}   \\ 
        GU Sgr	&	18:24:15.58	&	-24:15:26.35	& W & 0.66 & Mild    &	HdC3	C$_2$3 & Max.  {\tiny (23/07/11)} & {\small L27, H59}  \\ %
        HD 175893	&	18:58:47.29	&	-29:30:18.07	& W & 0.45 & Cold    &	HdC4	C$_2$3.5		& Max. {\tiny (16/07/10)}  & {\small P96a, B53}  \\ %
        IRAS1813.5-2419	&	18:16:39.20	&	-24:18:33.40	& W & 3.95 &  Cold    &	HdC6	C$_2$5	H0.5	& Max.  {\tiny (22/07/22)} &  {\small G07, T13}  \\ 
        MACHO 135.27132.51	&	18:19:33.76	&	-28:35:58.07	&   &	1.6	&  Cold    &				&  & {\small Z05}  \\ %
        MACHO 301.45783.9	&	18:32:18.60	&	-13:10:49.6	&   &	4.85	&  Cold    &				&  & {\small Z05}  \\ %
        MACHO 308.38099.66	&	18:19:27.42	&	-21:24:08.20	& W & 4.59  &  Cold    & HdC7 C$_2$5.5 &  Max.   {\tiny (13/04/22)}& {\small Z05}   \\ 
        MACHO 401.48170.2237	&	17:57:59.02	&	-28:18:13.1	& W & 3.65  &  Cold    &	HdC7	C$_2$5.5	Li:	& Max.  {\tiny (22/07/22)} & {\small Z05}  \\ 
        NSV11154	&	18:37:51.26	&	47:23:23.47	& T  & 0.21	&  Cold    &	HdC5:	C$_2$5.5		& Max.  {\tiny (22/06/19)} & {\small K11}   \\ %
        OGLE-GC-RCB-1	&	17:35:18.11	&	-26:53:49.24	&  & 4.09 &  Cold    &  &  & {\small T11}   \\ 
        R CrB	&	15:48:34.41	&	28:09:24.30	&  T & 0.07	& Warm   &	HdC1:	C$_2$2	H0.5	& Max.  {\tiny (21/06/19)} & {\small P97, E90, B35}   \\ %
        RS Tel	&	18:18:51.22	&	-46:32:53.43	& W & 0.24 & Mild   &	HdC3	C$_2$3		& Max.   {\tiny (16/07/10)}& {\small B53}  \\ %
        RT Nor	&	16:24:18.67	&	-59:20:38.57	& W & 0.63 &  Mild  &	HdC2	C$_2$3	H0.5	& Max.  {\tiny (14/03/22)} & {\small M51, W67}  \\ 
        RY Sgr	&	19:16:32.77	&	-33:31:20.33	& W & 0.41 & Warm  &	HdC1	C$_2$1	H1.5	& Max.  {\tiny (15/07/10)} & {\small P96b, B53, D65}  \\ %
        RZ Nor	&	16:32:41.66	&	-53:15:33.2	& W & 1.59 &  Mild  &	HdC2	C$_2$2.5	Li	& Max. {\tiny (14/03/22)}  & {\small F73, G78}  \\ %
        S Aps	&	15:09:24.54	&	-72:03:45.18	& W & 0.36 &  Cold    &	HdC5	C$_2$5	Li:	& Max.  {\tiny (17/07/10)}   & {\small P96c, W67}  \\ %
        SU Tau	&	05:49:03.73	&	19:04:21.89	&  T &	1.6	& Mild   &	HdC2: C$_2$- & Max. {\tiny (22/11/20)} & {\small B16, B53}  \\ %
        SV Sge	&	19:08:11.77	&	17:37:41.17	& W & 1.49  &  Cold    &	HdC6	C$_2$5.5		& Max.  {\tiny (24/07/11)}   &  {\small B53}  \\ %
        U Aqr	&	22:03:19.70	&	-16:37:35.28	& W & 0.10 &  Cold    &	HdC5	C$_2$5	Sr	& Max.  {\tiny (15/07/10)}   & {\small B79a, B79b}  \\ %
        UV Cas	&	23:02:14.66	&	59:36:36.69	&  T & 2.42	& Warm   &	HdC2:	C$_2$0.5		& Max. {\tiny (26/06/19)}   & {\small D14, P60, W67}  \\ %
        UW Cen	&	12:43:17.19	&	-54:31:40.76	&  & 1.03 & Warm   &		&  & {\small W67}   \\ 
        UX Ant	&	10:57:09.06	&	-37:23:55.13	& W & 0.25 & Warm   &	HdC1	C$_2$2	H1	& Max. {\tiny (16/02/22)}  & {\small E40, K90}  \\ %
        V CrA	&	18:47:32.31	&	-38:09:32.32	&  & 0.30 &  Mild  &		&  & {\small W67}  \\ 
        V1157 Sgr	&	19:45:42.36	&	-26:35:26.54	& W & 0.52 &  Cold    &	HdC6: C$_2$5.5		& $\Delta\sim$0.5 {\tiny (24/07/12)} & {\small L91}  \\ 
        V1783 Sgr	&	18:04:49.74	&	-32:43:13.63	& W & 1.55 &  Cold    &	HdC5	C$_2$5		& Max.   {\tiny (16/07/10)}  & {\small L91}  \\ %
        V2331 Sgr	&	17:23:14.55	&	-22:52:06.32	&  & 2.23 &  Cold    &		&  & {\small unpublished}  \\ 
        V2552 Oph	&	18:17:06.85	&	-23:57:51.35	& W & 2.19 & Warm    &	HdC1	C$_2$0.5	H1	& Max.  {\tiny (15/07/22)}   & {\small H02, R03}  \\ %
        V3795 Sgr$^{Std}$ &	18:13:23.56	&	-25:46:40.89 & W & 2.71 & Warm    &	HdC0	C$_2$0.5		& Max.	{\tiny (16/07/10)}   & {\small H72}	\\%
        V391 Sct	&	18:28:06.61	&	-15:54:44.13	& W & 5.19 &  Warm  &	HdC1	C$_2$0.5	H1 Li:	& Max.  {\tiny (15/07/10)} & {\small T13}   \\ %
        V4017 Sgr	&	18:18:02.29	&	-29:29:33.04	& W & 0.78 &  Cold    &	HdC4	C$_2$3		& Max.   {\tiny (31/07/12)}  & {\small H75}  \\ 
        V482 Cyg	&	19:59:42.57	&	33:59:27.94	&  T & 2.68	&  Warm  &	HdC1:	C$_2$1.5  & Max.   {\tiny (26/06/19)}  & {\small W49, W67}  \\ 
        V517 Oph$^{Std}$	&	17:15:19.73	&	-29:05:37.40	& W & 1.67 &  Cold    &	HdC6 C$_2$5.5 Li:	& Max.  {\tiny (15/07/22)}	& {\small K92}	\\
        V532 Oph	&	17:32:42.61	&	-21:51:40.76	& T  &	2.6	& Warm   &	HdC1: C$_2$1	&  Max.  {\tiny (02/08/22)} & {\small C09}  \\ %
        V739 Sgr	&	18:13:10.56	&	-30:16:14.73	& W & 1.45 &  Cold    &  HdC6: C$_2$5.5		& $\Delta\sim$0.4 {\tiny (16/07/10)} & {\small L91}  \\ 
        V854 Cen	&	14:34:49.40	&	-39:33:19.28	& W & 0.31 & Warm    &	HdC1	C$_2$0.5	H3b	& Max.  {\tiny (23/07/11)}  & {\small M86, K89, L89}  \\ 
        VZ Sgr$^{Std}$	&	18:15:08.57	&	-29:42:29.50	& W &  1.04 & Warm    &	HdC1	C$_2$0.5	H1	& Max. {\tiny (17/07/10)}    & {\small W67} \\ 
        WISE J004822.34+741757.4	&	00:48:22.34	&	74:17:57.4	&   &	1.16	& Cold   &				&  & {\small K21}   \\ 
        WISE J005128.08+645651.7	&	00:51:28.08	&	64:56:51.7	& T  &	3.46	& Hot/Warm   &See Section~\ref{sec:j005}	& Max. {\tiny (21/11/20)}   & {\small K21}   \\ 
        WISE J110008.77-600303.6	&	11:00:08.77	&	-60:03:03.6	& W & 5.45  &  Cold    &     HdC5:	C$_2$5	H2.5	& Unknown   {\tiny (07/06/12)}   & {\small T20}   \\ 
        WISE J132354.47-673720.8	&	13:23:54.47	&	-67:37:20.8	& W & 1.47  &  Cold    &	HdC6:	C$_2$5.5	Li:	& Max.   {\tiny (22/07/22)}  & {\small T20}   \\ 
        WISE J150104.50-563325.1	&	15:01:04.50	&	-56:33:25.1	& W & 5.40  &  Cold    &	HdC7: C$_2$- Li: & Unknown   {\tiny (24/07/12)}  & {\small T20}   \\ %
        WISE J160205.48-552741.6	&	16:02:05.48	&	-55:27:41.6	& W & 4.12  &  Cold    &	HdC5:	C$_2$5		& Unknown   {\tiny (07/06/12)}  & {\small T20}   \\ %
        WISE J161156.23-575527.1	&	16:11:56.23	&	-57:55:27.1	& W & 0.99  &  Cold    &	HdC5:	C$_2$5		& Unknown   {\tiny (07/06/12)}  & {\small S19, T20}   \\ %
        WISE J163450.35-380218.5	&	16:34:50.35	&	-38:02:18.5	&  & 1.75  &  Cold    &		&  & {\small T20}    \\ 
        WISE J164704.67-470817.8	&	16:47:04.67	&	-47:08:17.8	& W & $>$6  & Warm    &	HdC2: C$_2$- H0.5	& Unknown  {\tiny (24/07/13)}   & {\small T20}    \\ %
        WISE J170343.87-385126.6	&	17:03:43.87	&	-38:51:26.6	& W & $>$6  &  Cold    &	HdC5: C$_2$-	& Unknown  {\tiny (07/06/12)}   & {\small T20}    \\ %
        WISE J170552.81-163416.5	&	17:05:52.81	&	-16:34:16.5	&   &	1.27	&  Mild/Cold  &				&  & {\small K21}   \\ 
        WISE J171815.36-341339.9	&	17:18:15.36	&	-34:13:39.9	&  & $>$6  &  Cold    &		&  & {\small T20}    \\ 
        WISE J171908.50-435044.6	&	17:19:08.50	&	-43:50:44.6	&  & 5.17  &  Cold    &		&  & {\small T20}    \\ 
        WISE J172447.52-290418.6	&	17:24:47.52	&	-29:04:18.6	& W & 3.57  & Warm    &	HdC1	C$_2$0.5	H1	& Max.  {\tiny (01/08/12)}   & {\small T20}   \\ %
        WISE J172553.80-312421.1	&	17:25:53.80	&	-31:24:21.1	&  & $>$6  & Warm    &		&  & {\small T20}   \\ 
        WISE J172951.80-101715.9	&	17:29:51.80	&	-10:17:15.9	& W & 2.08  & Warm    &	HdC0	C$_2$0	H1 Li:	& Max.  {\tiny (24/09/2021)}   & {\small T20}   \\ %
        WISE J173202.75-432906.1	&	17:32:02.75	&	-43:29:06.1	&  & 1.23  &  Cold    &		&  & {\small T20}   \\ 
        WISE J173553.02-364104.3	&	17:35:53.02	&	-36:41:04.3	& W & 4.75 &  Cold    &	HdC6:	C$_2$5.5		& Unknown  {\tiny (24/07/13)}   & {\small T20}   \\ %
        WISE J173737.07-072828.1	&	17:37:37.07	&	-07:28:28.1	&  & 2.77 & Cold    &		&  & {\small K21}   \\ 
        WISE J173819.81-203632.1	&	17:38:19.81	&	-20:36:32.1	& W & 2.94 &  Cold    &	HdC6	C$_2$5.5		& Max.  {\tiny (16/07/22)}   & {\small S19, T20}   \\ %
        WISE J174111.80-281955.3	&	17:41:11.80	&	-28:19:55.3	& W & $>$6 & Warm    &	HdC1: C$_2$- H0.5	& Max.  {\tiny (15/07/22)}   & {\small T20}   \\ 
        WISE J174119.57-250621.2	&	17:41:19.57	&	-25:06:21.2	& W & 4.65 &  Mild  &	HdC3	C$_2$2		& Max.  {\tiny (21/07/11)}  & {\small T20}   \\ 
        WISE J174138.87-161546.4	&	17:41:38.87	&	-16:15:46.4	& W & 1.49 &  Cold    &	HdC6:	C$_2$5.5	Li:	& Unknown  {\tiny (26/07/13)}  & {\small T20}   \\ %
        WISE J174257.19-362052.1	&	17:42:57.19	&	-36:20:52.1	&  & 3.31 &  Cold    &		&  & {\small S19, T20}   \\ 
        WISE J174328.50-375029.0	&	17:43:28.50	&	-37:50:29.0	& W & 2.10 &  Cold    &	HdC6	C$_2$5.5	Li:	& Max.  {\tiny (02/08/12)}	& {\small S19, T20}	\\ 
        WISE J174645.90-250314.1	&	17:46:45.90	&	-25:03:14.1	& W & $>$6  & Mild    &	HdC3:	C$_2$3		& Max.  {\tiny (18/07/22)}  & {\small T20}   \\ %
        WISE J174851.29-330617.0	&	17:48:51.29	&	-33:06:17.0	& W & 4.07 & Mild   &	HdC2	C$_2$2.5		& Max.   {\tiny (03/08/12)} & {\small T20}   \\ %
        WISE J175031.70-233945.7	&	17:50:31.70	&	-23:39:45.7	& W & $>$6 &  Warm  &	HdC2:	C$_2$2.5		& Unknown  {\tiny (25/07/12)}  & {\small S19, T20}   \\ %
        WISE J175107.12-242357.3	&	17:51:07.12	&	-24:23:57.3	& W & 7.10 & Warm   &	HdC2: C$_2$-	& Unknown  {\tiny (25/07/12)}  & {\small T20}   \\ %
        WISE J175521.75-281131.2	&	17:55:21.75	&	-28:11:31.2	&   & 6.17 &   Cold   &			&  & {\small T20}   \\ 
        WISE J175558.51-164744.3	&	17:55:58.51	&	-16:47:44.3	& W & 3.46 & Warm   & HdC1: C$_2$0.5 & $\Delta\sim$0.5 {\tiny (16/07/22)} & {\small T20}   \\ 
        WISE J175749.76-075314.9	&	17:57:49.76	&	-07:53:14.9	& W & 5.20 &  Cold    &	HdC4:	C$_2$5		& Unknown  {\tiny (02/08/12)}  & {\small T20}   \\ %
        WISE J175749.98-182522.8	&	17:57:49.98	&	-18:25:22.8	& W &  $>$6 &  Cold    &  HdC6: C$_2$-	& Max.  {\tiny (16/07/22)}  & {\small T20}   \\ 
        WISE J180550.49-151301.7	&	18:05:50.49	&	-15:13:01.7	& W & 4.83 & Warm   &	HdC2: C$_2$0.5 H0.5	& $\Delta\sim$0.5 {\tiny (16/07/22)} & {\small T20}   \\ 
        WISE J181252.50-233304.4	&	18:12:52.50	&	-23:33:04.4	& W & 5.09 &  Cold    &	HdC4	C$_2$4.5	H0.5	& Max.  {\tiny (16/03/2022)}  & {\small T20}   \\ %
        WISE J181538.25-203845.7	&	18:15:38.25	&	-20:38:45.7	& W & $>$6 &  Cold    & HdC5:	C$_2$5	H0.5	& Unknown  {\tiny (24/07/12)}  & {\small T20}   \\ 
        WISE J181836.38-181732.8	&	18:18:36.38	&	-18:17:32.8	&  &  $>$6 & Cold   &		&  & {\small K21}   \\ 
        WISE J182235.25-033213.2	&	18:22:35.25	&	-03:32:13.2	& W & $>$6 & Cold   &      HdC7: C$_2$-    & Max.  {\tiny (15/07/22)} & {\small T20}  \\ 
        WISE J182334.24-282957.1	&	18:23:34.24	&	-28:29:57.1	&  & 1.15 &  Cold    &	&	& {\small T20}	\\ 
        WISE J182723.38-200830.1$^{Std}$	&	18:27:23.38	&	-20:08:30.1	& W & 3.38 &  Cold    &	HdC7:	C$_2$5.5	Li:	& Unknown  {\tiny (15/08/13)}  & {\small T20}   \\					
        WISE J182943.83-190246.2	&	18:29:43.83	&	-19:02:46.2	& W & 1.86 &  Cold    &	HdC5: C$_2$5		& $\Delta\sim$0.5 {\tiny (24/09/21)} & {\small T20}   \\ 
        WISE J183649.54-113420.7	&	18:36:49.54	&	-11:34:20.7	& W & 4.09 &  Cold    &	HdC7:	C$_2$5.5		& Unknown  {\tiny (24/07/12)}  & {\small T20}   \\					
        WISE J184158.40-054819.2	&	18:41:58.40	&	-05:48:19.2	& W & 4.76 &  Cold    &	HdC4:	C$_2$4		& Unknown  {\tiny (01/08/12)}  & {\small T20}   \\					
        WISE J184246.26-125414.7	&	18:42:46.26	&	-12:54:14.7	& W & 3.49 &  Cold    &	HdC7:	C$_2$5.5	Li:	& Unknown  {\tiny (27/07/13)}  & {\small T20}   \\					
        WISE J185525.52-025145.7	&	18:55:25.52	&	-02:51:45.7	& W & 4.24 & Warm    &	HdC1	C$_2$0.5	H1.5	& Max.  {\tiny (01/08/12)}  & {\small T20}   \\					
        WISE J185726.40+134909.4    &	18:57:26.40	&	13:49:09.4 &  & 2.39 & Cold   &   &  & {\small K21} \\ 
        WISE J190813.12+042154.1	&	19:08:13.12	&	04:21:54.1	&   &	3.83 &  Cold  &	&  & {\small K21} \\ 
        WISE J192348.98+161433.7	&	19:23:48.98	&	16:14:33.7	&   &	$>$6	&  Cold  &				&  & {\small K21}   \\ 
        WISE J194218.38-203247.5	&	19:42:18.38	&	-20:32:47.5	& W & 0.26 &  Cold    &	HdC6	C$_2$5.5	Li:	& Max.  {\tiny (07/06/12)}  & {\small L15, T20} \\					
        WX CrA$^{Std}$	&	18:08:50.49	&	-37:19:43.21	& W & 0.43 &  Cold    &	HdC5:	C$_2$5		& $\Delta\sim$0.4  {\tiny (17/07/10)}  & {\small W67}	\\ 
        XX Cam	&	04:08:38.75	&	53:21:39.35	&  T &	1.19	& Warm   &	HdC2: C$_2$-	& Max. {\tiny (22/11/20)} & {\small B48, Y48}   \\ %
        Y Mus	&	13:05:48.20	&	-65:30:46.62	& W & 2.22 & Warm    &	HdC0	C$_2$0.5		& Max.  {\tiny (17/07/10)}  & {\small W67}   \\ %
        Z Umi	&	15:02:01.37	&	83:03:48.63	&  T & 0.41	&   Cold   &	HdC5:	C$_2$5.5		& Max.  {\tiny (21/06/19)} & {\small B94}   \\ %
         \hline
		  \multicolumn{9}{c}{} \\ 
		  \multicolumn{9}{c}{Magellanic RCB stars} \\ 
         \hline
        ASAS J050232-7218.9	&	05:02:32.26	&	-72:18:53.56	& W & 0.32 &  Cold    &	HdC4	C$_2$5	Li:	& Max.  {\tiny (18/02/22)}  & {\small R79, O14}   \\				
        EROS2-LMC-RCB-1	&	05:14:40.22	&	-69:58:39.97	&    &  0.28	&  Cold    &		&  & {\small T09}   \\ 
        EROS2-LMC-RCB-2	&	05:10:28.53	&	-69:47:04.50	&   D &   0.27	&  Cold    &	HdC4	C$_2$5		& Max.    {\tiny (14/02/08)}  & {\small M03, T09}  \\			
        EROS2-LMC-RCB-3	&	04:59:35.83	&	-68:24:44.81	&   D &   0.35	&  Cold    &	HdC4:	C$_2$5	Li: Sr	& $\Delta\sim$0.4  {\tiny (13/02/08)}  & {\small R79, T09}   \\ 
        EROS2-LMC-RCB-4	&	05:39:36.98	&	-71:55:46.79	&    &  0.27	&  Cold    &		&  & T09   \\ 
        EROS2-LMC-RCB-5	&	06:04:05.48	&	-72:51:22.78	&   D &  0.43	&  Cold    &	HdC5	C$_2$5.5		& Max.  {\tiny (13/02/08)}  & {\small M03, T09}   \\ %
        EROS2-LMC-RCB-6	&	06:12:10.51	&	-74:05:09.94	&  & 0.47 &  Cold    &		&  & {\small T09}   \\ 
        EROS2-SMC-RCB-1	&	00:37:47.10	&	-73:39:02.40	&   &  0.10	&  Cold    &	&  & {\small M03, T04, T09}   \\ 
        EROS2-SMC-RCB-2	&	00:48:22.95	&	-73:41:04.65	&  & 0.10 &  Cold    &		&  & {\small T04, T09}   \\ %
        EROS2-SMC-RCB-3	&	00:57:18.15	&	-72:42:35.20	&   & 0.20	&  Cold    &		&	& {\small T04, T09}	\\ 
        HV 12842	&	05:45:02.88	&	-64:24:22.72	&   &	0.18	& Warm    &				&  & {\small P71, F72}  \\ %
        HV 5637	&	05:11:31.38	&	-67:55:50.81	&   &	0.32	&  Cold    &				&  & {\small P71, F72}   \\ %
        WISE J005010.67-694357.7	&	00:50:10.67	&	-69:43:57.7	& W & 0.05 & Warm    &	HdC2	C$_2$2		& Max.  {\tiny (25/08/21)}  & {\small T20}   \\ 
        WISE J005113.58-731036.3	&	00:51:13.58	&	-73:10:36.3	&  & 0.20 &  Cold    &		&  & {\small T20}   \\ 
        MSX-LMC-1795	&	05:42:21.90	&	-69:02:59.1	&   &	0.72	&  Cold    &	&  & {\small S09, M14, T20}   \\ 
        WISE  J055643.56-715510.7	&	05:56:43.56	&	-71:55:10.7	& W & 0.34 & Mild    &	HdC3	C$_2$4		& Max.   {\tiny (18/08/13)} & {\small T20}   \\ %
        MACHO-12.10803.56	&	05:46:47.74	&	-70:38:13.43	&   &	0.41	&  Cold    &				&  & {\small A01}   \\ %
        MACHO-16.5641.22	&	05:14:46.20	&	-67:55:47.64	&   &	0.3	&  Cold    &				&  & {\small BW83, A01}   \\ %
        MACHO-18.3325.148	&	05:01:00.38	&	-69:03:43.13	&   &	0.32	&  Cold    &				&  & {\small A01}   \\ %
        MACHO-6.6575.13	&	05:20:48.21	&	-70:12:12.54	&   &	0.37	&  ??$^{*}$  &				&  & {\small A01}   \\ %
        MACHO-6.6696.60	&	05:21:47.97	&	-70:09:57.09	&   &	0.27	& Mild    &				&  & {\small A01}   \\ %
        MACHO-79.5743.15	&	05:15:51.79	&	-69:10:08.64	&   &	0.23	&  Cold    &				&  & {\small A01}   \\ %
        MACHO-80.6956.207	&	05:22:57.38	&	-68:58:18.89	& W & 0.53 &  Cold    &	HdC6:	C$_2$5.5	Li:	& Unknown  {\tiny (19/02/22)}  & {\small A01}   \\ %
        MACHO-80.7559.28	&	05:26:33.92	&	-69:07:33.32	&   &	0.28	&  Cold    &				&  & {\small A01}   \\ %
        MACHO-81.8394.1358	&	05:32:13.35	&	-69:55:57.67	& W & 0.30 &  Cold    &	HdC6:	C$_2$5.5	H0.5 Li:	& Unknown  {\tiny (19/02/22)}  & {\small A96}   \\ 
        MSX-SMC-014	&	00:46:16.33	&	-74:11:13.60	&   &	0.1	&  Cold    &				&  & {\small K05}   \\ %
        W Men	&	05:26:24.53	&	-71:11:11.80	& W & 0.29 & Warm   &	HdC1:	C$_2$0.5	H0.5	&$\Delta\sim$0.3  {\tiny (15/07/22)}   & {\small F56, R70, F72}   \\ 
        KDM 5651	&	05:41:23.50	&	-70:58:01.8	& W & 0.37 &  Cold    &	HdC4	C$_2$5		& Max.  {\tiny (12/02/22)} & {\small M03, S09, T20}   \\ 
         \hline
		  \multicolumn{9}{c}{} \\ %
         \multicolumn{9}{c}{ Galactic dLHdC stars} \\ %
         \hline 
        A166$^{Pec}$	&	13:03:13.562	&	-19:09:22.679	&  &  0.22 &  Cold    &				&   & {\small T22}	\\			
        A182	&	15:47:39.0	&	-44:35:13.82	& W &  0.76 & Mild    & HdC2 C$_2$3 H0.5		&  Max.  {\tiny (09/04/21)}  &  {\small T22}   \\ 
        A183	&	15:50:33.036	&	-39:44:11.288	& W & 1.65  & Mild    & HdC3	C$_2$3 H1.5 Li:	&  Max.  {\tiny (20/05/21)}  &  {\small T22}   \\ %
        A223$^{Std}$	&	18:56:35.4	&	-16:09:12	& W & 1.60 & Mild     &	HdC3	C$_2$3 H0.5	&  Max.  {\tiny (14/04/21)}	&  {\small T22}	\\ 
        A226	&	19:02:26.53	&	-22:28:45.12	& W & 0.48 & Mild    &	HdC2	C$_2$3 H0.5	&  Max.  {\tiny (20/05/21)}  &  {\small T22}   \\ 
        A249	&	20:39:33.9	&	-51:23:01	& W & 0.08 & Warm    &	HdC1	C$_2$3 Sr H1	&  Max.  {\tiny (14/04/21)}  &  {\small T22}   \\ %
        A770	&	17:02:44.56	&	-33:35:15.493	& W & 1.67 &  Cold    &	HdC4	C$_2$5 H0.5 Li:	&  Max.  {\tiny (24/09/21)}  &  {\small T22}   \\ %
        A811	&	18:54:17.25	&	-11:50:23.16	& W & 2.19 & Mild     &	HdC3	C$_2$3.5 H1.5 Li:	&  Max.  {\tiny (25/08/21)}  &  {\small T22}   \\ %
        A814	&	18:59:28.34	&	11:49:33.84	& W & 3.05 & Mild     &	HdC4	C$_2$4 H0.5 Li: CN-	&  Max.  {\tiny (26/09/21)}  &  {\small T22}   \\ %
        A977	&	17:42:42.29	&	-24:18:05.78	& W & 3.68 &  Cold    &	HdC4	C$_2$4.5 Li:	&  Max.  {\tiny (18/10/21)}  &  {\small T22}   \\ %
        A980	&	18:11:35.62	&	01:54:32.59	& W & 1.15 & Warm    &	HdC0	C$_2$0.5 H1 Li:	&  Max.  {\tiny (18/10/21)}   &  {\small T22}   \\ %
        B42	&	18:13:40.5	&	-31:29:53	$^{Std}$ & W &  1.17 &  Cold    &	HdC4	C$_2$4.5 H0.5	&  Max.  {\tiny (11/04/21)}	&  {\small T22}	\\ 
        B563	&	17:19:29.03	&	-24:28:33.88	& W & 2.94 & Mild     &	HdC3	C$_2$4.5 H0.5 Li:	&  Max.  {\tiny (24/09/21)}  &  {\small T22}   \\ %
        B564	&	17:34:26.19	&	-41:05:22.17	& W & 1.82 & Warm    &	HdC0	C$_2$0.5 H0.5 Li:	&  Max.  {\tiny (10/09/21)}  &  {\small T22}   \\ %
        B565	&	17:40:48.48	&	-20:32:20.15	& W & 2.23 & Mild    &	HdC2	C$_2$3.5 H1 Li:	&  Max.  {\tiny (10/09/21)}  &  {\small T22}   \\ %
        B566	&	17:42:23.744	&	-21:32:00.735	& W & 2.19 & Mild    &	HdC2	C$_2$3 H0.5	& Max.  {\tiny (10/09/21)}  &   {\small T22}   \\ 
        B567	&	17:43:05.65	&	-22:24:57.42	& W & 2.90 & Warm    &	HdC1	C$_2$0.5 bl		&  Max.  {\tiny (10/09/21)}  &  {\small T22}   \\ 
        C17	&	17:05:40.13	&	-26:16:56.66	& W & 1.15  & Mild    &	HdC2	C$_2$3 H1.5 Li:	&  Max.  {\tiny (31/05/21)}  &  {\small T22}   \\ %
        C20	&	20:19:42.79	&	05:04:04.04	& W & 0.33 & Mild     &	HdC3	C$_2$3.5 H1.5 Li:	&  Max.  {\tiny (22/06/21)}  &  {\small T22}   \\ %
        C27	&	18:01:32.19	&	-38:56:32.30	& W & 0.61  & Mild    &	HdC3	C$_2$4	&  Max.  {\tiny (31/05/21)}  &  {\small T22}   \\ 
        C38	&	18:11:43.32	&	-28:14:21.31	& W & 1.41 & Mild     &	HdC3	C$_2$3 H0.5 Li:	&  Max.  {\tiny (21/06/21)}  &  {\small T22}   \\ %
        C526	&	13:56:40.21	&	-57:40:01.28	& W & 1.67 & Warm    &	HdC1	C$_2$0.5 H1	&  Max.  {\tiny (25/08/21)}  &  {\small T22}   \\ %
        C528	&	17:00:47.62	&	-33:30:55.81	& W & 1.74 & Mild     &	HdC3	C$_2$0.5 H2 Li:	&  Max. {\tiny (10/09/21)}  &  {\small T22}   \\ %
        C539	&	18:44:41.49	&	-14:15:02.21	& W & 2.97 & Mild     &	HdC3	C$_2$3 Sr	Ba &  Max.  {\tiny (24/09/21)} &  {\small T22}   \\ 
        C542	&	18:57:19.95	&	10:59:21.61	& W & 2.60 & Mild     &	HdC3	C$_2$2.5 H1.5 Li:	&  Max. {\tiny (26/09/21)}  &  {\small T22}   \\ 
        F152	&	05:15:28.99	&	-66:48:37.42	& W & 0.45 & Warm    &	HdC1	C$_2$1 H2e Li:	bl &  Max.  {\tiny (25/08/21)} &  {\small T22}   \\ 
        F75	&	19:44:40.59	&	-34:54:42.33	& W & 0.55 & Warm    &	HdC1	C$_2$0.5 H0.5 Li:	&  Max. {\tiny (21/06/21)}  &  {\small T22}   \\ %
        HD 137613	&	15:27:48.32	&	-25:10:10.13	& W & 0.37 &  Cold    &	HdC4	C$_2$3.5		&  Max.  {\tiny (17/07/10)} & {\small F92, B53}  \\ %
        HD 148839	&	16:35:45.79	&	-67:07:36.69	& W & 0.23 & Mild    &	HdC2	C$_2$3 H2 Li:	&  Max. {\tiny (17/07/10)}  & {\small W63}  \\ %
        HD 173409$^{Std}$	&	18:46:26.63	&	-31:20:32.08	& W & 0.45 & Mild    &	HdC2	C$_2$3 &  Max. {\tiny (16/07/10)}	& {\small P96a, B53}	\\%
        HD 182040	&	19:23:10.08	&	-10:42:11.54	& W & 0.59 & Mild    &	HdC2	C$_2$4		&  Max. {\tiny (16/07/10)}	& {\small P12, B53}  \\ %
        HE 1015-2050	&	10:17:34.23	&	-21:05:13.88	& W & 0.11 & Mild   &	HdC4	C$_2$5 Sr H1 Li: CN-	&  Max. {\tiny (07/06/12)}	& {\small G10}  \\ %
        M38	&	14:29:55.0	&	-57:27:55.7	& W & 4.60  &  Cold    &	HdC4	C$_2$4 H0.5	&  Max. {\tiny (13/02/22)}  & {\small unpublished}   \\ %
        P12	&	16:32:47.0	&	-43:49:13	& W & 4.20 & Mild     &	HdC3	C$_2$3.5 H0.5	&  Max. {\tiny (15/02/22)}  & {\small unpublished}   \\ %
\hline
\multicolumn{9}{l}{$^{Std}$: HdC stars used as standard for the definition of the main 8 classes} \\ 
\multicolumn{9}{l}{$^{Pec}$: HdC stars presenting a peculiar spectrum} \\ 
\multicolumn{9}{l}{$^{*}$: MACHO 6.6575.13 has yet to be observed spectroscopically in a bright phase} \\
\multicolumn{9}{l}{Spectrograph instruments: W: 2.3m/WiFeS, T: TNG/Dolores, D: 2.3m/DBS, see Table~\ref{tab:setups}} \\ %
\multicolumn{9}{l}{A$_V$: See Section~\ref{sec:obs}, Temperature group: See Section~\ref{subsec:mkprocess}} \\
\multicolumn{9}{l}{ References of the discovery papers:} \\ 
\multicolumn{9}{l}{ P97: \citet{1797RSPT...87..133P}, E90: \citet{1890MNRAS..51...11E}, F92: \citet{1892AstAp..11..765F}, P96a: \citet{1896ApJ.....4..142P},} \\
\multicolumn{9}{l}{  P96b: \citet{1896ApJ.....4..138P}, P96c: \citet{1896ApJ.....3..296P}, P12: \citet{1912ApJ....35..125P}, D14: \citet{1914AN....198..271D},} \\
\multicolumn{9}{l}{B16: \citet{1916BHarO.617....1B}, W24: \citet{1924BHarO.807....4W}, L27: \citet{1927BHarO.852....4L}, B35: \citet{1935ApJ....81..369B}, E40: \citet{1940BHarO.913....1E}, } \\
\multicolumn{9}{l}{B48: \citet{1948ApJ...107..413B}, Y48: \citet{1948ApJ...107..413Y}, W49: \citet{1949ApJ...109..538W},  M51: \citet{1951BHarO.920...32M}, B53: \citet{1953ApJ...117...25B},}\\
\multicolumn{9}{l}{ F56: \citet{1956MNRAS.116..583F}, H59: \citet{1959AJ.....64..241H}, P60: \citet{1960JO.....43...79P}, W63: \citet{1963MNRAS.126...61W}, D65: \citet{1965MNRAS.130..199D}, }\\
\multicolumn{9}{l}{  W67: \citet{Warner1967_hdcs}, R70: \citet{1970Obs....90..197R},  P71: \citet{1971SCoA...13.....P}, F72: \citet{1972MNRAS.158P..11F}, H72: \citet{1972IBVS..617....1H}, }\\
\multicolumn{9}{l}{ F73: \citet{1973MNRAS.161..293F}, H75: \citet{Hoffleit1975_discovery}, G78: \citet{1978MNRAS.185...23G}, B79a: \citet{1979ApJ...233..205B}, B79b: \citet{1979IAUS...83..305B},}\\
\multicolumn{9}{l}{   R79: \citet{1979ApJ...230..724R}, T81: \citet{1981ATsir1169....5T}, BW83: \citet{1983MNRAS.202P..31B},
M86: \citet{1986IAUC.4233....3M}}\\
\multicolumn{9}{l}{ K89: \citet{1989MNRAS.238P...1K}, L89: \citet{1989MNRAS.240..689L}, K90: \citet{1990Obs...110...90K}, L91: \citet{1991Obs...111..244L},}\\
\multicolumn{9}{l}{  F92: \citet{1992QJRAS..33..111F}, K92: \citet{1992Obs...112..158K}, B94: \citet{1994AJ....108..247B}, A96: \citet{1996ApJ...470..583A}, }\\
\multicolumn{9}{l}{  A01: \citet{2001ApJ...554..298A}, C02: \citet{2002PASP..114..846C}, H02: \citet{2002AAS...201.1711H}, M03: \citet{2003MNRAS.344..325M},}\\
\multicolumn{9}{l}{  R03: \citet{RaoLambert2003}, T04: \citet{2004AA...424..245T}, G07: \citet{2007PZ.....27....7G},  K05: \citet{2005ApJ...631L.147K},}\\
\multicolumn{9}{l}{  Z05: \citet{2005AJ....130.2293Z}, T08: \citet{2008AA...481..673T}, C09: \citet{2009PASP..121..461C}, S09: \citet{2009AcA....59..335S},}\\
\multicolumn{9}{l}{   T09: \citet{Tisserand2009_eroscloudstars}, G10: \citet{2010ApJ...723L.238G}, K11: \citet{2011PASP..123.1149K}, T11: \citet{Tisserand2011_oglestars},}\\
\multicolumn{9}{l}{  M12: \citet{2012ApJ...755...98M}, T13: \citet{Tisserand2013_asasdiscovery}, M14: \citet{2014MNRAS.439.1472M}, O14: \citet{2014JAVSO..42...13O},  L15: \citet{2015AA...575A...2L}, }\\
\multicolumn{9}{l}{ T20: \citet{Tisserand2020_plethora}, S19: \citet{2019MNRAS.483.4470S}, K21: \citet{2021ApJ...910..132K}, T22: \citet{Tisserand2022_dlhdcdiscovery}} \\
\hline

\end{longtable}
\end{landscape}
\twocolumn


\section{Classification Criteria}
\label{sec:criteria}

HdC stars are characterized predominantly by the extreme weakness and often complete absence of hydrogen in their spectra, and therefore do not exhibit strong Balmer lines, the Balmer jump, or the CH band. Stars colder than HdC type 1 show strong C$_2$ bands which dominate the blue end of the spectrum, and show CN bands in the red for stars colder than HdC type 2. Compared to typical carbon stars, HdC stars most resemble the C-N stars, as they have enhanced \textit{s}-process elements and weak $^{13}$C isotopic bands ($^{12}$C/$^{13}$C $\gtrapprox$ 100 \citep{Clayton2012_review}), compared to C-N stars with $^{12}$C/$^{13}$C $\sim$ 30-70 \citep{Lambert1986_hdcreview}. HdC stars also tend to have low Fe abundances \citep{Asplund2000}, and are found in low metallicity populations such as the thick disk, the Galactic bulge, and the Magellanic Clouds, especially the LMC bar \citep{Tisserand2020_plethora}.
The absolute magnitudes of HdCs in the SMC and LMC indicate that these stars have the same luminosity as supergiants, even while having a mass of only $\sim$\,0.8 M$_{\sun}$. 

In most stellar classification schemes, commonly used features for both temperature and luminosity classification include the hydrogen lines, many \textit{s}-process lines, and metallic lines. Care must be taken not to rely on the line strengths and ratios from these other schemes as HdC stars are nontraditional in all of these features, having no hydrogen lines, enhanced \textit{s}-process elements, and low metallicity. Metallic lines can still be used, however they must be compared directly to other heavy metallic lines that are minimally affected by nucleosynthesis within the star, e.g. Cr, Mn, Ti, etc. As the HdCs have a large range in the amount of \textit{s}-processing in their atmosphere, the \textit{s}-process spectral features are not of much use in the classification scheme.

Below we list each class and a non-exhaustive list of some important spectral features that can be used to help identify that class:

\begin{itemize}
\item HdC0:
\begin{itemize}
\item Continuum peaks in far blue ($\sim$ 4000 \text{\AA})
\item Presence of He I lines in the blue
\item Mn I 4030 \text{\AA} much stronger than Fe I 4046 \text{\AA}
\item Presence of C II lines
\item Strong Si II lines at 6347 and 6371 \text{\AA} (lines here are strong but there is more total flux in the continuum so they may appear to be weaker than in cooler stars)
\item N I feature blends with redward end of 8662 \text{\AA} line of Ca II triplet
\end{itemize}

\item HdC1:
\begin{itemize}
\item Mn I 4030 \text{\AA} stronger than Fe I 4046 \text{\AA}
\item Peak strength of C I lines, especially 6013, 6455 \text{\AA}
\item Strong O I 7004 \text{\AA} line
\item C I blend at 7115 \text{\AA} strong
\item Strong O I triplet at 7774 \text{\AA}
\end{itemize}

\item HdC2:
\begin{itemize}
\item Fe I 4046 \text{\AA} now equal to Mn I 4030 \text{\AA}
\item Ca I 4226 \text{\AA} becomes more prominent compared to neighboring lines, comes out near the bandhead
\item Ba II 4554 \text{\AA} begins to blend with CN bandheads
\end{itemize}

\item HdC3:
\begin{itemize}
\item Fe I 4046 \text{\AA} and 4271 \text{\AA} lines reach their peak depth
\item O I 7004 \text{\AA} line begins to dissipate
\end{itemize}

\item HdC4:
\begin{itemize}
\item Ti II 4444 \text{\AA} line begins to weaken but stays visible throughout the rest of the types (note typically used as a luminosity indicator)
\item Mg II 4481 \text{\AA} begins to weaken
\item depending on the C$_2$ band strengths, tends to be where C$_2$ bands around 6000 \text{\AA} become easy to spot (note no $^{13}$C$^{12}$C bands!)
\end{itemize}

\item HdC5:
\begin{itemize}
\item CN bands becoming easier to spot
\end{itemize}

\item HdC6:
\begin{itemize}
\item The C$_2$ bands begin to reach 0 flux at their deepest point
\item Metallic spectrum becoming quite rich
\end{itemize}

\item HdC7:
\begin{itemize}
\item Continuum peaks in far red (> 9000 \text{\AA})
\item Weak-lined metallic spectrum still weaker than C, N, O features
\item Very few individual lines able to be identified, most features are blended
\item C$_2$ bands bluer than 4500 \text{\AA} have very little flux
\end{itemize}
\end{itemize}

\begin{figure*}
	\includegraphics[width=\textwidth]{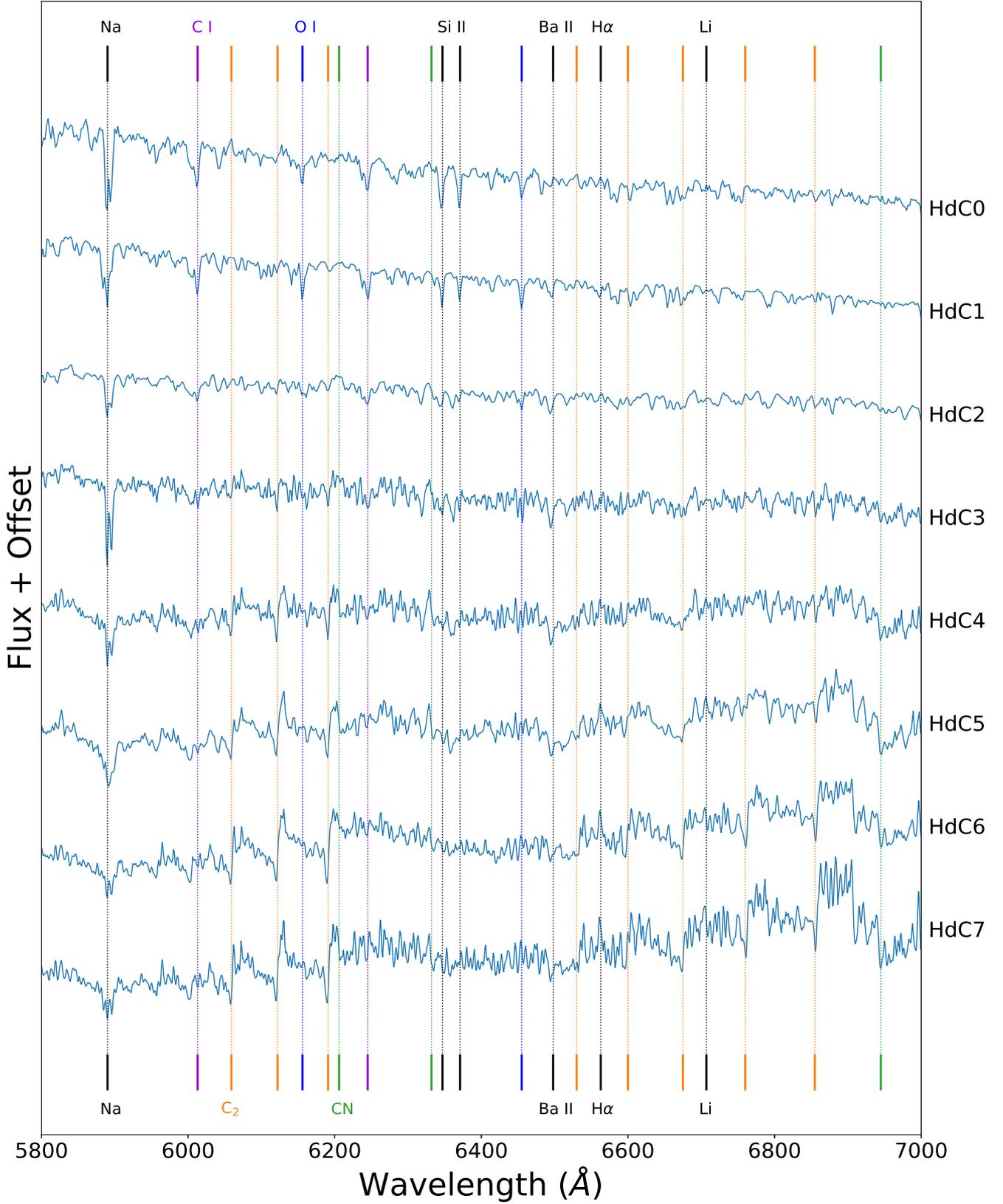}
    \caption{The spectral region of 5800 \text{\AA} to 7000 \text{\AA} for each of the HdC standard stars. The HdC0 class at the top indicates the warmest group of stars, and HdC7 at the bottom indicates the coolest group of stars. Each spectrum is divided by its flux at 6900 \text{\AA}} and offset from each other. The vertical lines at the top indicate the features most commonly found in the warm stars, and those at the bottom are features best seen in cool stars. All purple lines indicate C I features, blue lines indicate O I, orange lines indicate C$_2$ bandheads, and green lines indicate CN bandheads. Black lines are labeled directly with the element they represent.
    \label{fig:sequence_zoom}
\end{figure*}

The spectra of the cooler classes are significantly more dense than the warmer classes, leading to fewer unblended identified lines, as most of their features are blends of metallic and band features. In order to classify a new HdC, we recommend beginning by finding the few classes which best match the continuum of the new star, and then overlaying the spectra and viewing the features closely. A particularly enlightening region of the spectrum is the region from 5800 \text{\AA} to 7000 \text{\AA}, shown in Figure~\ref{fig:sequence_zoom}. In this region the classifier can easily view multiple C$_2$ bands (confirming the weakness of the $^{12}$C$^{13}$C bands at 6100 and 6168 \text{\AA}), the first CN bands at 6215 and 6910 \text{\AA}, and many important neutral atomic features such as C I, O I, and Si II. 

Predominant features in the HdC spectrum also include the Ca II features--- namely the H \& K lines and the IR triplet--- and the Na I D lines. These features seem to vary between stars even within the same class by a significant amount. Na I D is known to vary with gas along the line of sight \citep{Blondin2009_naDline}, which is problematic for our stars as they tend to lie at large distances from the Earth, and some HdCs have significant foreground dust \citep{Tisserand2022_dlhdcdiscovery}. Ca II, another prominent feature in the HdC spectra, is sensitive to both metallicity and luminosity. We further discuss the Ca II triplet in Section~\ref{subsec:luminosity_calib}. We do not include Na I D or the Ca II features in our classification criteria, nor do we include peculiarities regarding the depth of these features.

\subsection{The Carbon Index}
\label{sec:carbon_index}

As mentioned in Section~\ref{subsec:final}, we based the carbon index for HdCs on the carbon index for typical carbon stars. The HdC stars show a large number of both C$_2$ and CN bands \citep{1983MNRAS.202P..31B}, however this index refers specifically to the C$_2$ bands. The CN bands are highly correlated in strength, except in a few cases, which we note for those stars specifically (see Section~\ref{sec:misc_features}). Using this carbon index, C$_2$0 indicates no visible C$_2$ bandheads, and C$_2$6 indicates the strongest bands in our sample. Even the warmest HdC type, HdC0, has stars with weak but clearly visible C$_2$ bandheads at 4737 and 5165 \text{\AA}. At carbon index C$_2$3.5 is the first indication of the C$_2$ bandhead at 4383 \text{\AA}, most clearly visible in classes HdC3 and HdC4 (Figures~\ref{fig:set3} and \ref{fig:set4}). Thus, the presence of the 4383 \text{\AA} band is a useful checkpoint in determining the carbon index. 

As we have many partial spectra in the sample, there are also quite a few stars for which we do not have enough wavelength coverage to view the C$_2$ Swan bands in the blue. For these stars, we do not indicate a C$_2$ index, opting to use either C$_2$- or fully omitting the carbon index. 

\subsection{The Hydrogen Index}
\label{sec:hydrogen_index}

The majority of HdC stars exhibit no detectable 4300 \text{\AA} CH band or Balmer lines. However, in 47 out of the 128 stars with spectra, we find the Balmer lines in absorption, and in emission in one star, F152. We measured the pseudo-equivalent widths (pEWs)\footnote{Note that we refer to these measurements as pseudo-equivalent widths rather than simply equivalent widths as HdCs do not generally show a true continuum, and therefore we can only measure from the pseudo-continuum.} of the H$\alpha$ line and rounded to the nearest 0.5 \text{\AA} to create the hydrogen index, which we denote using H\#. 
Of the stars with detectable hydrogen lines, 25 out of the 48 are dLHdCs, and 23 out of the 48 are RCBs. This translates to 75.8\% of the total number of dLHdCs and 24.2\% of the total number of RCBs exhibiting H$\alpha$ in their spectrum, confirming that dLHdC stars are more likely to show hydrogen lines than their dust-producing counterparts \citep{Tisserand2022_dlhdcdiscovery}. A very active RCB star known to have quite strong Balmer lines, V854 Cen, has a hydrogen index of H3, and as it is the only star which has a visible Balmer discontinuity, we additionally included a b to the notation, making its hydrogen index ``H3b''. Of the stars with detectable H, none of the stars in our sample, even those with measurable Balmer lines, had clearly detectable CH bands.

\subsection{The Lithium Index}
\label{sec:lithium_index}

Lithium is an element only previously observed in four RCBs (R CrB, UW Cen, RZ Nor, and SU Tau) until it was potentially observed in a number of dLHdCs \citep{Tisserand2022_dlhdcdiscovery}. We find detectable features at the location of the 6707 \text{\AA} lithium line in 35 HdC stars--roughly 1/3 of the total sample--17 of which are dLHdCs (50.0\% of dLHdC sample) and 18 of which are RCBs (18.9\% of RCB sample). This feature is quite weak in all  cases, and lies within a strong C$_2$ band in the cooler stars, making its identification as lithium questionable. Nonetheless, we indicate these stars via a ``Li:'' in the classification, in the hopes that future high-resolution follow-up will confirm or deny the identification of this line. As for the four stars with known surface lithium, only RZ Nor decisively shows a Li feature. Thus we denote a ``Li'' in its class, indicating a clear identification. R CrB and SU Tau's spectra are taken at too low of resolution to confirm the lithium line, as they were obtained with our TNG/Dolores setup (see Table~\ref{tab:setups}), whereas UW Cen does not currently have a maximum light spectrum for us to view. High resolution follow up of Li-enriched stars will give us a better idea of how much lithium enhancement is typical, and will guide future HdC evolutionary models (such as those in \citet{Munson2021}, \citet{Crawford2020_mesa}, \citet{Schwab2019}, \citet{Lauer2019}, \citet{Zhang2012_mergermodels}, and \citet{Longland2012_lithium}). While most of these models do not consider lithium abundances, \citet{Longland2012_lithium} show that surface lithium is possible by careful consideration of the merger parameters and the initial $^{3}$He in the model. This result was corroborated by \citet{Zhang2012_mergermodels}. Future observations of our potentially Li-enriched stars will help solidify the amount of surface lithium necessary in HdC models.

\subsection{Miscellaneous Peculiar Features}
\label{sec:misc_features}

The set of Sr-rich stars found by \citet{Crawford2022_strontium} are denoted by ``Sr'' in their classification. The star C539 additionally has a ``Ba'' to denote its extraordinarily strong Ba II lines at 6140 and 6496 \text{\AA} compared to the rest of its class. The star A166 is not included in the current classification, however it would also be considered a Ba-rich star. There are a range of known Sr and Ba abundances in the HdC stars, however we do not have the resolution to comment on the relative abundances of these elements with our spectral classification scheme.

The isotope $^{13}$C is known to be extraordinarily weak in HdCs, and also serves as the likely source of neutrons for the \textit{s}-process via the $^{13}$C($\alpha$,n)$^{16}$O reaction. The lack of $^{13}$C is most obvious by the absence of the $^{12}$C$^{13}$C bands at 6100 and 6168 \text{\AA} and also at the 4744 \text{\AA} band. Identification of this isotope is more difficult in the warmer stars as the C$_2$ Swan bands are much weaker, however all HdCs with known strong $^{13}$C are warm stars -- V CrA, V854 Cen, VZ Sgr, and UX Ant have measured $^{12}$C/$^{13}$C < 25 \citep{Rao2008_vcra,Hema2012_carbon}. Our spectra are not of high enough resolution to see $^{13}$C in these warm stars. We do not find any cool HdCs with detectable $^{13}$C at the 4744 \text{\AA} band, however with higher resolution one would be able to place a limit on the actual $^{13}$C abundance.

HdC stars are known to be enhanced in nitrogen \citep{Asplund2000}, however they are also known to have weak CN bands compared to other carbon stars \citep{Morgan2003_notmkmorgan}. The CN bands both increase in strength with an increase in luminosity, and decrease in strength with a decrease in metallicity \citep{Keenan1993_carbonclass}.
Since the HdC stars not only have a (small) range of luminosities (see Section~\ref{subsec:luminosity_calib}) but also a range of low metallicities and a range of nitrogen abundances, it is therefore not surprising the strength of CN shows some variance within the same class and carbon index. In our system, we denote stars with especially weak CN bands by ``CN-'' and those with especially strong bands are denoted via ``CN+''.

\citet{Asplund2000} discovered a small class of four stars which are denoted ``minority'' stars based on their differences in abundances from the ``majority'' class of RCBs in their sample. These minority stars are VZ Sgr, V3795 Sgr, V854 Cen, and V CrA. While the minority class is a diverse class, they are predominantly characterized by low metallicity and extreme abundance ratios, especially S/Fe and Si/Fe. Upon further inspection into the abundances of these stars, the ratios S/Fe and Si/Fe are large not because of S or Si enhancements, but because of weak Fe. While we can clearly identify the Si II lines at 6347 and 6371 \text{\AA}, we do not have high enough resolution to estimate an abundance for Fe, and therefore we cannot distinguish minority stars from majority stars.

Two stars, F152 and B567, are denoted with a ``bl'' for ``broad-lined'' to indicate that the absorption features are particularly broad. Both stars are dLHdCs, and F152 additionally has an H$\alpha$ feature in emission, and has some indication of a cold dust shell around it \citep{Tisserand2022_dlhdcdiscovery}. These stars are thus particularly interesting as they shows many features that RCBs would in a decline phase, despite being classified as dustless. 

In addition to individual feature peculiarities, we find 22 stars with a continuum peculiarity. These are EROS2-CG-RCB-1, SU Tau, UV Cas, V739 Sgr, XX Cam, WISE J110008.77-600303.6, WISE J172447.52-290418.6, WISE J174111.80-281955.3, WISE J174645.90-250314.1, WISE J175031.70-233945.7, WISE J175558.51-164744.3, WISE J175749.98-182522.8, WISE J180550.49-151301.7, WISE J181538.25-203845.7, WISE J182235.25-033213.2, WISE J005010.67-694357.7, B565, B567, C542, M37, P12, and HE 1015-2050. This type of peculiarity can have multiple sources, and perhaps the most likely is an incorrect A$_V$ value or an instrumental effect. As mentioned in Section~\ref{sec:obs}, we found the best possible A$_V$ in the literature, however for stars in dense stellar regions there is no guarantee this accurately represents the true amount of foreground dust. Additionally, if the star had a recent decline the dust surrounding the star may also affect and distort the continuum. If there are no decline features in the spectrum (see \citet{Clayton1996_review} and \citet{Rao2004_UWCendecline} for an overview of RCB decline spectra) 
and no photometric data for the observation date, it is not straightforward to untangle a faulty dereddening from a decline. However, for each of the continuum peculiar stars, their individual spectral features do not differ strongly from their assigned classes if you ignore the continuum effect. We chose not to include these in the classification directly as it is not clear whether these are intrinsic or extrinsic effects.


\section{Calibration}
\label{sec:calib}

\subsection{Color Index and Temperature}
\label{subsec:color_calib}

\begin{figure}
	\includegraphics[width=\columnwidth]{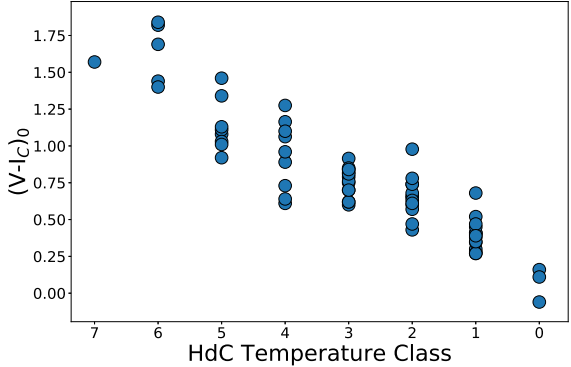}
    \caption{The intrinsic V-I$_C$ colors versus the HdC temperature class for our sample.}
    \label{fig:colors}
\end{figure}

Color calibration of the spectral classes can be particularly useful when direct temperature estimation is difficult, such as with the HdC stars. Figure~\ref{fig:colors} shows the range of intrinsic V-I$_c$ (hereafter simply V-I) color for HdC stars in each temperature class. 
These colors were originally collected for \cite{Tisserand2022_dlhdcdiscovery} using a combination of the AAVSO, ASAS-SN, and Pan-STARRS databases \citep{ASASSN1,ASASSN2,panSTARRS}. Estimating the intrinsic V-I colors for the RCB stars is especially difficult due to both their deep and long episodic declines and their maximum light oscillations. RCB star pulsations have a larger V band amplitude than the dLHdC stars, ranging from 0.2 - 0.4 mag for the RCBs and 0.05 - 0.15 mag for the dLHdCs, with some dLHdC stars not exhibiting a detectable level of variability from the ASAS-SN survey \citep{Tisserand2022_dlhdcdiscovery}. These pulsations indicate variation in the photospheric temperature and thus colors of these stars as well. From monitoring the colors of the eponymous R CrB during its 1988 decline and subsequent phase of maximum light, \cite{Cottrell1990_rcrbdecline1988} shows that R CrB's pulsations of 0.38 mag in V coincide with a change in V-I color of 0.145 mag. Additionally, the V-I color increases dramatically upon the star entering a decline phase. Therefore, it is paramount that the values used for V and I magnitudes be measured while the star is at maximum light, and if possible, at the same pulsational phase. However, due to the nature of large scale photometric sky surveys, coincident V and I mag estimates are not guaranteed. Additionally, the intrinsic V-I for these stars is dependent on the assumed V band extinction. As described in Section~\ref{sec:obs}, we take care to find the best possible estimate of A$_V$ for each star, however they are still subject to errors. 
As seen in Figure~\ref{fig:colors}, each temperature class has a fairly small range of V-I, listed in Table~\ref{tab:color_calib}, forming a clear relationship between HdC class and V-I color. Most stars in each class span a range of 0.1 to 0.2 mag, roughly the same as the variation in V-I during a typical RCB pulsation. 

There are two ways in which we can go about estimating temperature ranges for the HdC classes. One is to use the temperature estimates we compiled from the HdC literature (see Table~\ref{tab:lit_temperatures}), and the other is to calibrate to the colors of typical MK dwarfs. We present the temperature ranges found by both of these methods in Table~\ref{tab:color_calib}.

\begin{figure*}
	\includegraphics[width=\textwidth]{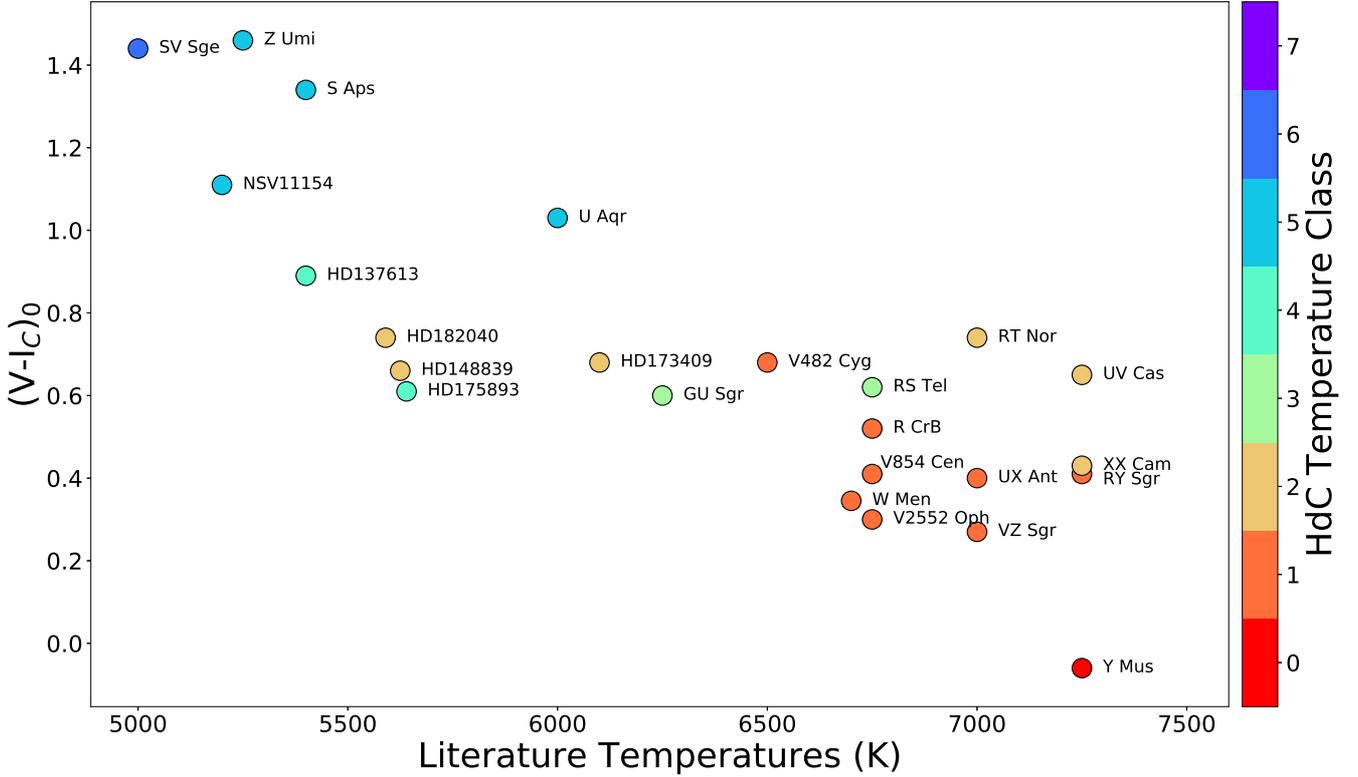}
    \caption{The intrinsic V-I$_C$ colors versus the those stars with temperatures listed in the literature (see Table~\ref{tab:lit_temperatures}). The HdC temperature class is indicated via the colorbar on the right. Each star is labelled. Note that not every star with a listed temperature from the literature has a measured V-I$_C$ color.}
    \label{fig:colortemps}
\end{figure*}

For 24 HdC stars, we can use the estimates of their temperatures from various literature (Table~\ref{tab:lit_temperatures}) to compare with their V-I colors and assigned temperature classes, as shown in Figure~\ref{fig:colortemps}. Here we see a strong correlation between V-I and temperature estimates, however we also notice tension between the assigned classes and the temperatures-- i.e. the HdC assigned class does not vary smoothly with temperature. We are not sure of the exact cause of this tension, but it is likely due to both discrepancies in the temperature calculations and different methods of temperature estimation. As an example, the three stars most clearly affected by this are RT Nor, UV Cas, and XX Cam. All three of these temperature estimates are from the same work \citep{Asplund2000}, which the majority of HdC temperatures are from, and therefore should be consistent with the other stars. While the temperature classes for XX Cam and UV Cas are estimates made from low-resolution red spectra, our spectrum for RT Nor should be a good representation of the true spectral appearance at maximum light. Without our own estimation of the temperature of these stars, it's not clear why some stars do not follow the expected trend. While we expect some scatter due to errors in V-I estimates and temperature estimates, we do not currently understand all the potential sources of errors in these measurements. Additionally, we have very few temperature estimates for the coolest HdC stars, and the HdC6 and HdC0 classes only have one star with an estimated temperature. Thus, it is extremely difficult to use the temperatures we have compiled from the literature as a method of temperature calibration. Nonetheless, we include the ranges of estimated temperatures for each class in Table~\ref{tab:color_calib}. 

In addition, we compared these V-I color ranges to those for typical MK dwarfs presented in \cite{Pecaut2013_mkcolorchart} to estimate a range in T\textsubscript{eff} for the HdC classes, listed in Table~\ref{tab:color_calib}. We emphasize that MK dwarfs and HdC supergiants have vastly different atmospheric properties, especially noting differences in metallicity, hydrogen content, carbon-to-oxygen ratio, and surface gravity (log\textit{g}). All of these factors, along with the effective temperature of the star, play an important role in the opacity, and thus directly affect the spectral colors \citep{Bell1989_colors,Bessell1998_colors,Huang2015_colors}.
Currently there is no comprehensive temperature calibration of V-I colors for other metal-poor supergiant stars in the literature. While we include the temperature ranges we have found using this method, we concede that this method is extremely likely to introduce a systematic offset in the estimated temperatures.

In Figure~\ref{fig:colors}, one can see that there are a few stars for which their V-I value is significantly different from that of the rest of their classmates. These stars do not appear to have different underlying continua when viewing their spectra, and since the V and I bandpasses are well represented in our spectral range, more investigation must be done to fully understand why these stars differ in V-I color.
Indeed, our sample is lacking in robust V-I estimates, especially for the coolest classes of stars. We are monitoring the maximum light V-I photometry of many HdCs\footnote{Our photometric monitoring tool for the HdC stars can be found at \href{https://dreams.anu.edu.au/monitoring/}{https://dreams.anu.edu.au/monitoring/}} in order to further calibrate the classification system. For the HdC6 and especially HdC7 classes, it is difficult to place an accurate V-I range due to the lack of data for these stars. Regardless, we still provide an estimate of the temperature based on the assumption that the HdC classes are in order of descending temperature in both the MK and literature calibrations.

Many of the typical line strengths and ratios used to gauge temperature in the MK and carbon star classification systems rely on either the hydrogen Balmer series lines or the iron lines, neither of which are easily used in HdC stars. While hydrogen lines are nearly absent and thus unusable, the HdC stars also have low metallicities and therefore have weak metallic spectra, making the iron lines difficult to measure at intermediate resolution. Higher resolution spectra are thus necessary to verify the temperature ranges of these stars. Nonetheless, the ratio of iron lines to other metallic lines minimally affected by nucleosynthesis such as chromium and calcium can be used to augment the HdC temperature calibration found from V-I colors, as is done in the traditional MK classification system, as long as one can satisfactorily estimate the pEWs of these lines. In this work, we base our temperature calibration principally on the V-I colors. 

The two different methods of calibrating the V-I colors to the surface temperature of HdCs give different overall temperature boundaries-- using MK dwarfs gives the HdCs a range from $\sim$3800 - 8500 K, whereas using temperatures from the literature gives a range from $\sim$4750 - 7250 K. To further explore these temperature boundaries, we can look to related classes of stars at the warm and cool ends of the HdC temperature distribution.

The class of DY Per like variables lie to the cold end of our HdC set. The cool star DY Per has a (V-I)$_0$ color of 2.28 mag. This is additionally $\sim$0.5 mag redder than the reddest HdC for which we have a V-I estimate. This implies an approximate temperature of $\sim$3500 K using the same aforementioned MK dwarf calibration. There is also a direct temperature estimate of DY Per from \citet{Keenan1997_dyper} of 3494 K, which is in agreement with the V-I color comparison to MK dwarfs. This implies that the coolest HdC stars must be warmer than $\sim$3500 K. For a detailed spectroscopic comparison of the cool HdCs to the DY Per variables and a discussion on the potential temperature differences between these two classes, see Section~\ref{subsec:dyper_calib}.

\begin{table}
    \centering
    \caption{HdC Color Calibration}
    \begin{tabular}{cccc}
         HdC & (V-I)$_0$ Range & T\textsubscript{eff} Range (K) & T\textsubscript{eff} Range (K) \\
         Sequence &  & (MK Dwarfs calib.) & (Literature calib.) \\
         \hline
         HdC0 & 0.1 - 0.2 & 7800 - 8500 & $\sim$7250 \\
         HdC1 & 0.25 - 0.4 & 6800 - 7800 & 6500 - 7250\\
         HdC2 & 0.45 - 0.65 & 5800 - 6800 & 5590 - 7250 \\
         HdC3 & 0.7 - 0.9 & 5000 - 5800 & 6250 - 6750 \\
         HdC4 & 1.0 - 1.25 & 4500 - 5000 & 5400 - 5640 \\
         HdC5 & 1.25 - 1.5 & 4100 - 4500 & 5200 - 6000 \\
         HdC6 & 1.6 - 1.75 & 3900 - 4100 & $\sim$5000 \\
         HdC7 & unknown & <3900 & <5000
    \end{tabular}
    \label{tab:color_calib}
\end{table}

\begin{figure*}
	\includegraphics[width=\textwidth]{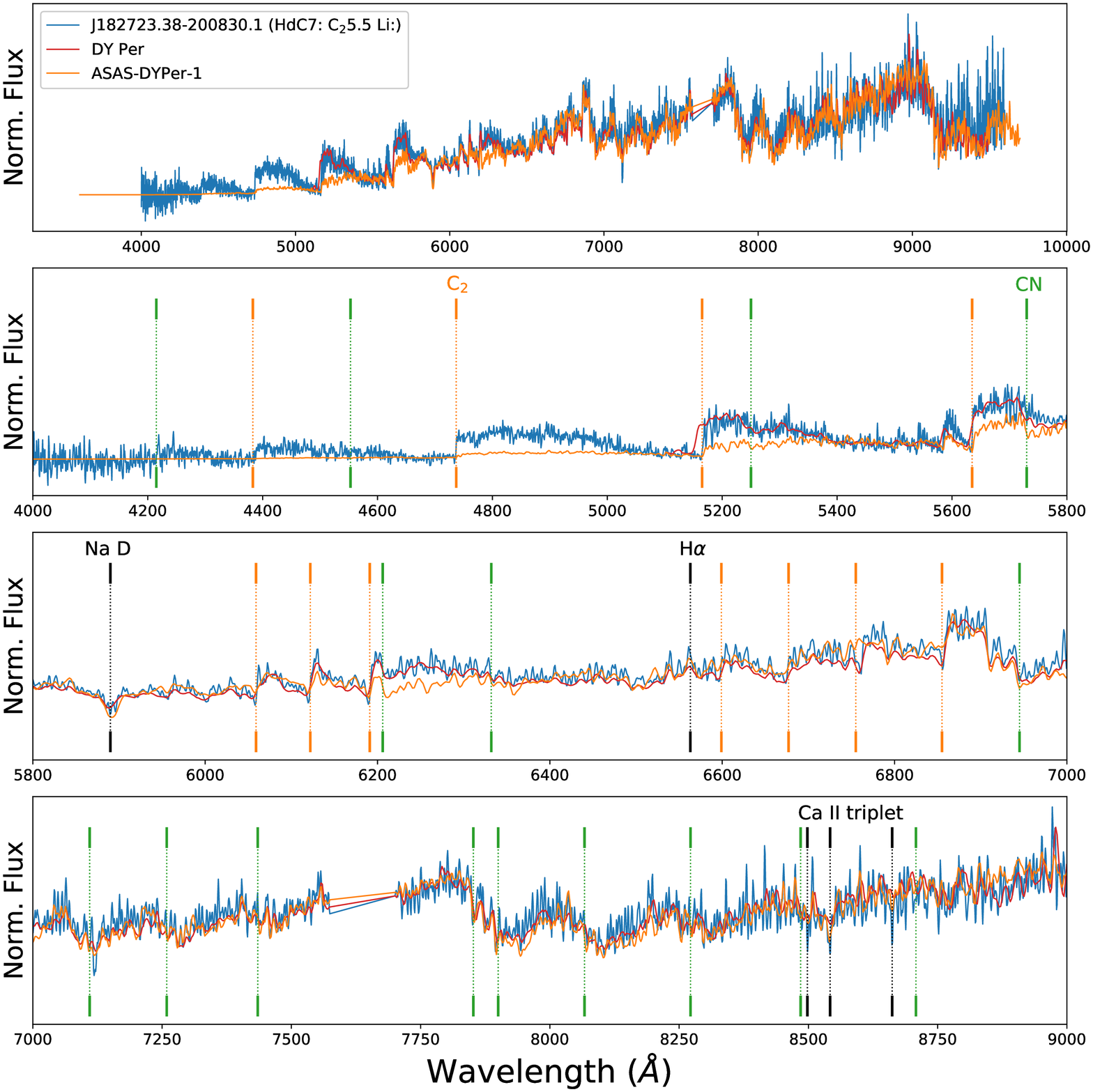}
    \caption{The spectrum of the HdC7 standard star, WISE J182723.38-200830.1 (in blue), overlaid with the spectrum of the eponymous DY Per (in red) and another DY Per type variable, ASAS-DYPer-1 (in orange). All spectra are normalized such that the region just bluer than 7000 \text{\AA} are at the same normalized flux. The top panel shows the full spectral range for all stars, where the three following panels show the 4000 to 5800 \text{\AA}, 5800 to 7000 \text{\AA}, and 7000 to 9000 \text{\AA} ranges. Note that the spectrum for DY Per is at a lower resolution than the other two spectra.We label the Ca II IR triplet, the rarely seen H$\alpha$, and Na D features in black, the C$_2$ bandheads in orange, and the CN bandheads in green.}
    \label{fig:dypercompare}
\end{figure*}

At the warm end of the HdC stars, we can compare to known Extreme Helium (EHe) stars, one of the coldest of which is FQ Aqr at T\textsubscript{eff} = 8750 $\pm$ 250 K \citep{Pandey2001_coolehe}. Using the Pan-STARRS colors and the calibrations of \citet{Tonry2012_panstarrs}, we can estimate a V-I color for FQ Aqr of $\sim$0.2 mag. Notice that this is actually redder than the warmest HdC star that we have a V-I estimate for. The color overlap of cool EHe stars with warm HdC stars can be seen clearly in Figure~10 of \citet{Tisserand2022_dlhdcdiscovery}. We can therefore assume that the warmest HdC stars should also have temperatures at or near $\sim$8750 K. 

When comparing the warm HdC stars to the cool EHes, one easily notices the strong appearance of the C II lines in EHe stars at this temperature (see Figure 1 of \citet{Pandey2001_coolehe}). The WiFeS spectra used in our classification set are not high enough resolution to resolve the C II lines in HdC spectra, however we can see the C II lines clearly in V3795 Sgr in published high resolution spectra (see Figure 1 in \citet{Lambert1994_merefacts}). We note that in typical oxygen-rich stars, C II lines are not seen in stars cooler than type B, however as HdCs and EHes are carbon-rich, the carbon features are therefore much stronger and seen in much cooler atmospheres. V3795 Sgr, an HdC0 C$_2$0.5 star, has clearly visible C II lines in published high resolution spectra, though not to the strength of those seen in FQ Aqr. However, V3795 Sgr also has visible C$_2$ bands as indicated by its carbon index, especially the 4670 and 5165 \text{\AA} bands which are  resolved even in our intermediate resolution spectra (see Figure~\ref{fig:set0} to view the C$_2$ bands for this star). The existence of both molecular carbon and ionized carbon in the same atmosphere is a rare observation, only possible for our warmest stars, even though the spectra in this work are not able to resolve the C II features directly and simultaneously.

In addition to C II lines, the HdC0 class also exhibits He I lines in the blue, much like the EHe stars. He I lines are seen as late as F0 supergiant stars near the Ca II H line, however in hydrogen-rich stars these lines are surrounded by the high energy Balmer transitions, to which the He I lines dwarf in comparison. In the HdC stars, the He I lines are much easier to spot as the Balmer transitions are not strongly present. Thus we infer that the HdC0 class can be approximately calibrated to the F0 supergiants, however, care must be taken to consider the effects of different abundances of elements such as H and He in the HdC stars when performing such a comparison.


\subsection{Luminosity}
\label{subsec:luminosity_calib}

As multiple HdC stars lie in the Magellanic clouds and many of the Galactic HdCs have parallaxes measured by the Gaia mission--thus having good distance estimates--we know that HdCs have supergiant luminosities \citep{Tisserand2009_eroscloudstars,Tisserand2022_dlhdcdiscovery}. The range of absolute V band magnitudes for HdC stars is between -2 and -5, and the Magellanic population of HdCs appear surprisingly brighter than Galactic counterparts at cool temperatures \citep{Tisserand2022_dlhdcdiscovery}. This would put the majority of HdC stars at a corresponding luminosity class ranging from II to Ia. Typical luminosity sensitive features used in other classification systems include the ratio of CN to C$_2$ bands and the ratio of Sr or Y lines to nearby Fe lines. Neither of these features are suited for use in HdC stars (see Section~\ref{sec:misc_features} and Section~\ref{sec:criteria}, respectively).
Additionally, the Ca II IR triplet is used in some cool star IR classification systems as an indicator of luminosity. Indeed this is a very prominent feature in HdC spectra, however it is highly dependent on the metallicity of the star and exhibits strong departures from local thermodynamic equilibrium (LTE) when comparing to model atmospheres \citep{Osorio2022_catrip}. In order to use the Ca IR triplet as a luminosity indicator, we would need to use some other proxy to control for the metallicity of the star. This is outside the scope of this work at the current time as we do not have robust metallicity estimates for very many stars.
Therefore we do not provide a luminosity classification for HdCs in this work, although we can confirm their luminosity is similar to traditional supergiants through their distances.


\section{Comparison to DY Per Type Variables}
\label{subsec:dyper_calib}

DY Persei, another rare carbon star variable, and the class of variable stars like it (DY Per type variables) have been theorized to be ``cool cousins'' of RCBs \citep{Alksnis1994_dypertheory,Bhowmick2018_coolcousins}, seeing as they have similar dust declines and are also hydrogen-deficient. The dust declines of these stars appear different to those of the RCBs, with the DY Per type declines being shallower and more symmetrical. While we cannot confirm or deny their relationship to the RCBs and the dLHdCs, we acknowledge that they lie under the HdC umbrella by definition, and thus we compare their spectra to our sample. We have spectra of 26 DY Per types, one from WiFeS, and the rest from our DBS setup, therefore we only have one spectrum spanning the entire spectral range 3800 to 9000 \text{\AA}. We additionally use the spectrum of the eponymous DY Per taken with our TNG/Dolores setup (see Table~\ref{tab:setups}). 

DY Per itself has been given a type of C-Hd4.5 C$_2$6 in \citet{Keenan1997_dyper}, and we show a comparison to our HdC7 standard, WISE J182723.38-200830.1 (HdC7: C$_2$5.5 Li:) and another DY Per type, ASAS-DYPer-1, in Figure~\ref{fig:dypercompare}. The spectra share a striking similarity, with WISE J182723.38-200830.1 and DY Per being the warmest at nearly the same temperature, ASAS-DYPer-1 being the coolest. Close inspection of the spectral range 5800 - 7000 \text{\AA} (shown in the third panel), one can truly see how similar these spectra are. There are only slight differences in the metallic spectrum, which could be the effect of the difference in temperature or an implicit difference in metallicity.
We additionally see that the CN bands at $\sim$5250 and 5730 \text{\AA} are stronger in DY Per than in the RCB stars. This is indeed true for all the DY Per types.

It seems that the prototype DY Per is approximately an HdC class warmer than the rest of the DY Per types that we have spectra of. All of the Magellanic EROS2 DY Per types \citep{Tisserand2009_eroscloudstars} are of approximately the same temperature, based on spectral comparison. The only DY Per like star which can be deconstructed using our PCA projection (i.e. the only DY Per type variable with a spectrum from 4500 to 9000 \text{\AA}) is ASAS-DYPer-1. Doing so places this star where one would infer a cooler group of HdCs would be within the PCA projection.

As we only have a full spectrum of one such star, we chose not to include the DY Per types in our classification system for now. It appears that they span a temperature range slightly cooler than the HdCs, and we refrain from assigning a carbon index until we can view more of the C$_2$ bands. However, from our present analysis we do not see any strong differences between the spectra of the coolest RCBs and the DY Per types other than the strength of the CN bands. The only other known difference between the stellar classes are the shapes and depths of their dust declines. The dust declines of these stars are enigmatic, and unfortunately it is beyond the scope of this paper to theorize possible differences between the dust production of RCBs and DY Per types.

\section{The Case of WISE J005128.08+645651.7}
\label{sec:j005}

During our analysis, we found one star, WISE J005128.08+645651.7, which appears to border between the warmest HdCs and the coolest EHe stars. This star is at a geometric distance of 5.34$^{+0.57}_{-0.53}$ kpc \citep{BailerJones2021_gaiadistances} with a visual extinction of A$_V$ $\sim$ 3.4 \citep{Green2019_extinction} and an apparent V mag of $\sim$13.55 mag \citep{ASASSN2}. This corresponds to an absolute V band magnitude of M$_V$ $\sim$ -3.5 $^{+0.25}_{-0.22}$ mag, which places the star at the faint end of the warm HdC brightness distribution, but similar to the brightest of the five hot RCB stars (See Fig. 10, \citet{Tisserand2022_dlhdcdiscovery}).

\begin{figure*}
	\includegraphics[width=\textwidth]{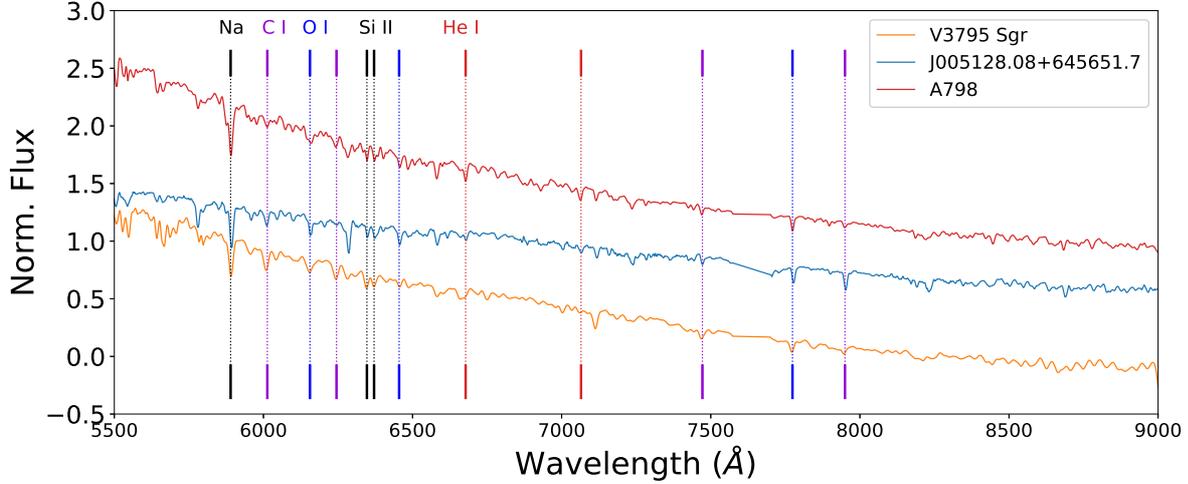}
    \caption{WISE J005128.08+645651.7 plotted in blue with the HdC0 standard star, V3795 Sgr in orange and an EHe star A798 in red. V3795 Sgr and A798 have been smoothed to the same resolution as WISE J005128.08+645651.7. Note that this star's continuum is slightly peculiar as well, which we find very common in the TNG/Dolores setup of spectra. We label a few of the most important spectral features-- Na D and Si II in black, the C I features in purple, the O I features in blue, and the He I features in red.}
    \label{fig:potentialehe}
\end{figure*}

In Figure~\ref{fig:potentialehe} we show the spectrum of this star with our HdC0 standard, V3795 Sgr, and the EHe star A798 discovered in \citet{Tisserand2022_dlhdcdiscovery}. As WISE J005128.08+645651.7 was observed using the TNG/Dolores setup, the spectrum is quite low resolution and only contains the red side. HdC0 and cool EHe stars such as A798 are in the range of late A/early F stars, which have nearly featureless red spectra. Therefore it is not straightforward to confirm that WISE J005128.08+645651.7 is an EHe without the blue side of the spectrum required to observe the He I lines. However, one can clearly see the resemblance between these three spectra, which are dominated by atomic and ionized lines.

This star was discovered by \citet{2021ApJ...910..132K}, who noted emission lines in the NIR spectrum similar to those observed in the hot RCB stars \citep{Demarco2002_hotrcbs}, which would also indicate a much warmer temperature than even the warmest HdC stars. However, this star does not show the optical emission lines seen in other hot RCBs. Additionally, the ASAS-SN survey \citep{ASASSN1,ASASSN2} observed a 1.5 magnitude decline at JD$\sim$2457500, and the star was discovered via the presence of a circumstellar dust shell in the infrared \citep{Tisserand2012_IRcolors,Tisserand2020_plethora}. Both of these are RCB-like properties. Thus, here we refrain from assigning it a class until we obtain further observations, however we are continuing to monitor this interesting star to probe the boundary between the EHe stars, the hot RCBs, and the warm HdC0 class.

\section{Conclusions and Final Discussion}
\label{sec:conclusions}

The HdC stars have traditionally not been included with other carbon stars both due to their rarity and their unique stellar composition. In this work, we have shown that the recent large influx of new RCBs and dLHdCs allows us to begin to classify these stars, and that many of the techniques used to classify carbon stars continue to be applicable to HdCs. There are, however, a number of spectral differences that one must keep in mind that prohibit the use of certain spectral features such as the Balmer series, the CH band, the CN bands, and the \textit{s}-process lines. Indeed, care must also be taken when considering the metallic spectrum of HdCs, as they are known to have low metallicity, and therefore heavy metallic lines must only be considered relative to each other. Nonetheless, we have created a spectral classification spanning the large temperature range of HdC stars which includes an index for the molecular carbon bands, as well as indices for H, Li, Sr, and CN. 

While we took special care to ensure our sample contained spectra taken while each star was at maximum light, we did not account for pulsational phase at the observation time. As mentioned in Section~\ref{subsec:color_calib}, a typical RCB pulsation ranges from 0.2 to 0.4 mag in V \citep{Cottrell1990_rcrbdecline1988,Lawson1997_rvpulsations}. 
These variations in V band magnitude have a corresponding change in intrinsic V-I color and indicate a change in the photospheric temperature. It is therefore likely that stars with especially large pulsation amplitudes such as RY Sgr may have different spectral appearances based on phase. In future works we hope to characterize the spectral variation during an oscillation cycle and ensure stability in a star's classification.

An overarching theme in recent HdC works has been in uncovering the difference between the dust-producing and dustless HdCs. Currently, the only known spectroscopic differences between RCBs and dLHdCs are that dLHdCs have stronger $^{18}$O isotopic bands in the infrared, higher likelihood to be Sr-rich or show Li lines, stronger H abundances, and weaker CN bands when compared to RCBs \citep{Karambelkar2022_oxygen18,Crawford2022_strontium,Tisserand2022_dlhdcdiscovery}. This is consistent with the results found in this work. In Figure~\ref{fig:histogram}, we show the relative occurrence of dLHdCs and RCBs in each class. With the small number of stars in each class, it is difficult to comment on the distribution of dLHdCs compared to RCBs, however it is clear that there are no dLHdCs in the five coldest temperature classes. This could be a result of dLHdCs forming as a result of slightly different WD-merger populations, as is explored in \citet{Tisserand2022_dlhdcdiscovery} and \citet{Karambelkar2022_oxygen18}, however it could also indicate that our current method of discovering these stars is not effective at finding cold dLHdCs, seeing as cold HdCs are fainter and thus more difficult to discover. We do not find any additional differences between the dLHdCs and RCBs in this work.

Additionally, we begin to systematically compare the RCBs and dLHdCs to the DY Per variables. We do not include the latter class in our system as we do not have many complete spectra, however we find that they are very similar to our coolest stars, with the only obvious difference being the cooler temperatures and the strength of the CN bands. It is not clear whether this difference in CN would be enough to explain the differences in the declines of these two stars, nor what would cause such a difference in CN strength. However, it seems likely that the stars are indeed related in some way, as their spectra are remarkably similar.  

In this work, we have classified all stars for which we have maximum light spectra (with the exception of two stars). This turns out to be 128 out of the 196 total stars
-- roughly 65\% of HdC stars classified. There are two stars deliberately excluded from classification and those are ASAS-RCB-6 and A166. A166 is a unique star in our sample, as it's been separated from the rest of the HdCs in every work it has appeared in since its discovery \citep{Tisserand2022_dlhdcdiscovery,Karambelkar2022_oxygen18,Crawford2022_strontium}. A summary of many of its known peculiarities can be found in Section 4.6 of \citet{Tisserand2022_dlhdcdiscovery}. ASAS-RCB-6 and A166 have extremely similar spectra and share many of the same unique properties, with the exception of the Ba strength. Indeed, their spectra do not fit well with any class of HdC stars. In addition to their peculiar continua, it is not straightforward to match their spectral features with any class. We are not sure why these stars are so different from other HdC stars, however we intend to follow up on them in high resolution and continue to study them. 

In the future, we intend to obtain intermediate resolution spectra for the stars not included in this study so that they too may be classified. Many of these stars were in dust declines during recent observational periods, or they may have been unobservable during the time frame that we typically use to observe HdC rich regions. In addition, this classification system has pointed out a number of interesting stars for high-resolution spectroscopic follow-up. Of particular interest are those stars with observable Li lines, as previously Li had only been observed in four RCB stars. Modeling the surface abundances of Li in HdCs has proven to be quite difficult, and having new observations of these stars may shed some light on how that element persists on the surface \citep{Crawford2020_mesa,Munson2021}. This new MK-like spectral classification system for the HdCs can now aid in the identification of these unique stars for follow up, and will help to advance our knowledge of HdC stars and their formation and evolution.

Lastly, we would like to comment on the overall distribution of HdC stars in each class. As seen in Figure~\ref{fig:histogram}, the classes HdC1 through HdC6 all have very nearly the same number of total stars, with HdC0 and HdC7 having fewer. It is not clear whether this has implications regarding the formation scenarios for these stars. As mentioned earlier, the cooler stars are fainter and thus more difficult to uncover. It is also possible that cooler stars may produce more dust, which could be studied by a thorough investigation of J-W2 colors. This would also make the cooler stars more difficult to locate and observe, diminishing their population size artificially.
As for the HdC0 class, we are also unsure why there appears to be fewer of these stars. In more recent works, we've seen that roughly 2/3 of the total RCBs are ``cool'' which in this case means they have either C$_2$ or CN bands present in their spectra \citep{Tisserand2020_plethora} (see Section~\ref{sec:methods}, note this study predates the use of the term ``mild''). 
In our classification system, the classes which would be considered cool are HdC4 through HdC7, where mild classes would be HdC2 and HdC3, and warm classes HdC0 and HdC1. Using this as a reference, our sample shows that there are 66 cool stars, 36 mild stars, and 26 warm stars out of the total 128 classified stars. Considering that the classes have roughly equal population size and that the cool stars span four classes, it is not surprising that they make up approximately half of the classified set. If you combine the cool and mild numbers, as would have been done in early HdC studies, then the stars with strongly visible molecular bands make up $\sim$80\% of the HdC stars.

This new classification system will help streamline discovery of unique HdC stars and help us to compare the HdCs with similar variable star classes in the future. We are following up on many of the unique stars in higher resolution, and continuing to amass intermediate resolution spectra of the stars that are not classified here. Additionally, we have shown that the DY Pers overlap with the coolest RCB stars in temperature and their only obvious difference is in the strength of their CN bands. In the future, we hope to bring these stars fully under the HdC umbrella if possible, and continue to broaden our understanding of this rich and unique set of stars.


\section*{Acknowlegements}

We thank the reviewer, Simon Jeffery, for his numerous helpful comments which greatly improved the work.

CC is grateful for support from National Science Foundation Award 1814967. PT acknowledges also financial support from ''Programme National de Physique Stellaire'' (PNPS) of CNRS/INSU, France. AJR was supported by the Australian Research Council through award number FT170100243. IRS was supported by the Australian Research Council through award number FT160100028. This article is based on observations made in the Observatorios de Canarias del IAC with the Galileo National Telescope (TNG) telescope operated on the island of La Palma by the Centro Galileo Galilei of INAF (Istituto Nazionale di Astrofisica) in the Observatorio del Roque de Los Muchachos.

\section*{Data Availability}
Table 4 of this work can be downloaded from CDS, along with the V and I band magnitudes, accepted SIMBAD identifiers, and all spectra presented.

\typeout{}
\bibliographystyle{mnras}
\bibliography{rcb_bib,rcb_table}

\bsp	
\label{lastpage}
\end{document}